\documentclass[a4paper,fleqn,usenatbib]{mnras}

\usepackage[T1]{fontenc}
\usepackage{ae,aecompl}

\usepackage{graphicx}	
\usepackage{amsmath}	
\usepackage{amssymb}	

\usepackage[ruled,norelsize]{algorithm2e}
\DeclareMathOperator{\cov}{\mbox{cov}}

\title{Towards the geometry of the universe from data}

\author[H. L. Bester et al.]{
Hertzog L. Bester,$^{1}$\thanks{E-mail: g07b1135@campus.ru.ac.za}
Julien Larena,$^{1}$
Nigel T. Bishop$^{1}$
\\
$^{1}$Department of Mathematics, Rhodes University, Grahamstown, 6140, South Africa}

\date{Accepted XXX. Received YYY; in original form ZZZ}

\pubyear{2015}

\begin{document}
\label{firstpage}
\pagerange{\pageref{firstpage}--\pageref{lastpage}}
\maketitle

\begin{abstract}
We present a new algorithm that can reconstruct the full distributions of metric components within the class of spherically symmetric dust universes that may include a cosmological constant. The algorithm is capable of confronting this class of solutions with arbitrary data and opens a new observational window to determine the value of the cosmological constant. In this work we use luminosity and age data to constrain the geometry of the universe up to a redshift of $z = 1.75$. We show that, although current data are perfectly compatible with homogeneous models of the universe, simple radially inhomogeneous void models that are sometimes used as alternative explanations for the apparent acceleration of the late time universe cannot yet be ruled out. In doing so we reconstruct the density of cold dark matter out to $z = 1.75$ and derive constraints on the metric components when the universe was 10.5 Gyr old within a comoving volume of approximately 1 Gpc$^{3}$.
\end{abstract}

\begin{keywords}
cosmology: observations -- dark energy -- dark matter
\end{keywords}

\section{Introduction}
Assuming that general relativity correctly describes gravity on cosmological scales the problem of reconstructing the cosmological metric can be posed as a characteristic initial value problem (CIVP). Spherically symmetric dust models that may include a cosmological constant (also known as $\Lambda$ Lema\^itre-Tolman-Bondi ($\Lambda$LTB) models) have, after all gauge freedoms have been fixed, two free functional degrees of freedom and a free parameter viz. the cosmological constant $\Lambda$. Ideally the two functional degrees of freedom should be fixed using data generated by astrophysical models that do not presuppose any specific cosmology. This is a major issue for the \emph{observational cosmology programme} \cite{1985PhR...124..315E,Stoeger92.0, Stoeger92.1, Stoeger92.2, Stoeger92.3, Stoeger94, Maartens96, Stoeger97, araujo00, Araujo2009zh, araujo09.1, araujo10, araujo11} with research pertaining to observations in non-symmetric universes ongoing \cite{Hellaby:2013ts,Hellaby:2000ef}. This paper however is not about observations in non-symmetric universes. Our aim instead is to specify an algorithm that can determine the full distribution of metric components in a spherically symmetric universe given observational data. As such we do not make any attempt to justify how the data are obtained. A detailed investigation into the model dependence of currently available data, as well as data expected from future surveys, has been relegated to an accompanying paper that will be released at a later stage. Note also that the algorithm we propose tests for radial inhomogeneities on large scales but does not address the averaging or backreaction problems (for reviews of these topics see \cite{Clarkson:2011zq,Buchert:2011sx} and references therein).\\
After verifying that the algorithm performs as expected on simulated data we investigate what constraints can be derived from currently available luminosity (from the Union 2.1 compilation \cite{Suzuki:2011hu}) and age (from Cosmic Chronometers \cite{Moresco:2012by}) data. Both data sets are reported as functions of redshift $z$ and we convert the luminosity data into angular diameter distance data while age data are used to get a measure of the longitudinal expansion rate. We have also used the best fit Planck value of $H_0 = 67.3 \pm 1.2 ~ \mbox{km} \mbox{s}^{-1} \mbox{Mpc}^{-1}$ \cite{Ade:2013sjv}. Thus we effectively assume that model independent data for $D(z)$ and $H_\|(z)$ are available without questioning the validity of such data in $\Lambda$LTB cosmologies. We will further go on to show how data for the energy density of cold dark matter $\rho(z)$ and the redshift drift $\frac{dz}{dw}(z)$ can be incorporated into the algorithm. Since only two free functions are required to set the initial conditions of the CIVP some care has to be taken to incorporate more data without over-specifying the model. Note that all the initial data except the value of the cosmological constant can in principle be set with any two of $D(z)$, $H_\|(z)$ and $\rho(z)$. In section~\ref{sec:DisMod} we derive an expression for the redshift drift and in section~\ref{sec:algorithm}, we also show how the value of $\Lambda$ can be inferred from redshift drift data. It has long been known that redshift drift data will be an invaluable observable in cosmology (see for example \cite{1962ApJ...136..319S,Loeb:1998bu,Corasaniti:2007bg,Geng:2015ara,2007MNRAS.382.1623B,PhysRevLett.100.191303}) but as far as we are aware this is a new result, at least for $\Lambda$LTB models. \\
The numerical integration scheme we employ to solve the CIVP requires continuous functions as input. Since observational data are reported at discrete values of the redshift we need a way to smooth data. Gaussian process regression (GPR) provides a convenient semi-parametric Bayesian methodology for this purpose. The main difficulty with using GPR, or in fact any non-parametric smoothing algorithm, is making sure that the reconstructed functions obey certain physical constraints. As we explain in section~\ref{sec:physcons}, there exists a relatively simple way to enforce all physical constraints if we choose a particular set to sample from viz. $H_\|(z), \rho(z)$ and $\Lambda$. Once these have been specified the model is completely determined by solving the Einstein field equations in observational coordinates. A likelihood can then be assigned to any set of initial conditions by comparing the solution to the observational data. This provides a means of performing inference on $H_\|(z), \rho(z)$ and $\Lambda$. However we stress from the onset that the algorithm we present requires redshift drift data to perform inference on $\Lambda$. Without $\frac{dz}{dw}(z)$ data the best we can do is marginalise over the value of $\Lambda$ by specifying some suitable prior. Prior specification is an important aspect of this algorithm that will be discussed in section~\ref{sec:algorithm}. \\
By now a number of algorithms have been proposed to solve the cosmological inverse problem (see \cite{Walters:2012ns,Redlich:2014gga,Sapone:2014nna,2014MNRAS.438L...6V} and references therein). This paper extends the formalism in \cite{Bester:2013fya}. To keep it as concise as possible we have omitted some details. In particular we refer the reader to that paper, and \cite{prd2012}, for further details regarding our formulation of the observational and CIVP formalism. \\
The paper is structured as follows: the next section highlights some key differences between the $\Lambda$LTB and FLRW models that we exploit to test for radial inhomogeneity. We then give a brief description of the integration scheme used to numerically solve the Einstein field equations in observational coordinates. To put our results within the standard cosmological context we also compute the coordinate transformation that relates observational coordinates to the standard comoving 1+3 cosmological coordinates. In section~\ref{sec:SmoothData} we give a brief overview of how we employ Gaussian process regression to smooth the data on the current past lightcone (PLC0) with particular emphasis on how we enforce physical constraints. In section~\ref{sec:algorithm} we present the algorithm that allows us to reconstruct the distributions of the metric functions inside the past lightcone (PLC). We then verify that the algorithm performs as expected on simulated data. Finally in section~\ref{sec:results} we present constraints from currently available data. We conclude with some prospects for future research.\\
For notational convenience we denote $1+z = u$ throughout. A subscript zero can refer to a quantity evaluated on either the PLC0 eg. $D(w_0,v) = D_0(v)$ or the current time slice eg. $H_{\perp}(t_0,r) = H_{\perp 0}(r)$, the meaning should be unambiguous from the context. The lower case Latin letter $f$ is used to refer to probability distribution functions. Note in this paper an overdot refers to a partial derivative w.r.t. the coordinate $w$ and a prime refers to a partial derivative w.r.t. the coordinate $v$ and we work in units where $c = G = 1$ throughout.

\section{Framework}
\subsection{The model}\label{sec:Model}
\begin{figure}
\includegraphics[width=0.45\textwidth]{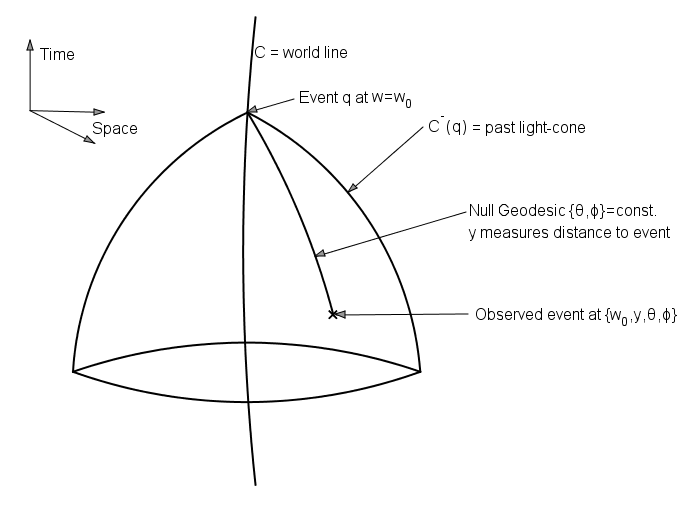}
\caption{A schematic illustration of the observational coordinate system. \label{fig:ObsCoords}}
\end{figure}

The $\Lambda$LTB cosmological model describes a spherically symmetric dust distribution \cite{Humphreys:1998tc} in a universe that may contain a cosmological constant \cite{Valkenburg:2011tm}. In spherical synchronous comoving coordinates $x^{\tilde{a}} = [t,r,\theta,\phi]$ (hereafter comoving coordinates or CC) the $\Lambda$LTB metric can be written as 
\begin{equation}
ds^2 = -dt^2 + X^2(t,r)dr^2 + R^2(t,r)d\Omega^2,
\label{LTBMet1}
\end{equation} 
where $d\Omega^2$ is the usual solid angle on the sphere, $r$ is a comoving radial coordinate and $t$ measures proper time in the frame of the observer. One of the first integrals of the EFE's allows us to relate $X$ to $R$ via $X = g(r)\partial_r R$ where $g(r)$ is an integration function that can be related to the space-time curvature. The EFE's can further be used to write down the analogue of the Friedmann equation in $\Lambda$LTB models as
\begin{equation}
H_\perp\left(R\right) = H_{\perp 0} \sqrt{\left(\frac{R}{R_0}\right)^{-3} \Omega_{m0} + \left(\frac{R}{R_0}\right)^{-2} \Omega_{K0} + \Omega_{\Lambda0}},
\label{LTBFried}
\end{equation}
where $H_{\perp} = \frac{\partial_t R}{R}$ is the transverse (or perpendicular to the line of sight) expansion rate and the dimensionless density functions now depend or $r$. Explicitly they are defined by
\begin{equation}
\Omega_{m} = \frac{\kappa \rho}{3 H_{\perp}^2}, \quad \Omega_{\Lambda} = \frac{\Lambda}{3 H_{\perp}^2} \quad \mbox{and} \quad \Omega_{K} = 1 - \Omega_{m} - \Omega_{\Lambda}, 
\end{equation}
where $\rho = \rho(t,r)$ is the energy density of cold dark matter and $\Lambda$ is the cosmological constant. The expression \eqref{LTBFried} can be used to get the age of the universe as
\begin{equation}
t - t_B(r) = \frac{1}{H_{\perp 0}}\int^{R}_0 \frac{dR^*}{R^*H_\perp(R^*)}. \label{gett0}
\end{equation}
where $t_B(r)$ is a function of integration (the \emph{bang time}) which can be scaled such that $t_B(0) = 0$. In particular we get the current age of the universe at our space-time location by evaluating \eqref{gett0} with $R = R_0$ along the central worldline of the observer (henceforth $\mathcal{C}$) \footnote{Evaluated in practise by transforming this elliptic integral into one of the Carlson symmetric forms}. This fact will be used to set the time grid for the numerical integration scheme. \\
Comoving coordinates are probably the most intuitive coordinate system for cosmology. However they are not the most natural coordinate system when it comes to interpreting data. For that we need observational coordinates $x^{a} = [w,v,\theta,\phi]$ (schematically illustrated in Figure~\ref{fig:ObsCoords}). The metric in these coordinates can be written as
\begin{equation}
ds^2 = -A(w,v)dw^2 + 2dwdv + D^2(w,v)d\Omega^2,
\label{ObsMet1}
\end{equation}
where $w$ is a coordinate such that $w = $~const. defines the PLC. Gauge fixing $w$ to measure proper time along the central worldline of the observer then implies that $A(w,0) = 1$. The coordinate $v$ measures the null affine distance down the PLC and, with this form of the metric, is necessarily non-comoving. The coordinates $\theta$ and $\phi$ are spherical coordinates and are the same in the two coordinate systems. These definitions ensure that there is no gauge freedom left in these coordinates. This is important because it means we can specify \emph{any} $\Lambda$LTB model by fixing two functional degrees of freedom and the parameter $\Lambda$. However it can be difficult to interpret a cosmological model in terms of these coordinates mainly because they entangle spatial and temporal evolution. Therefore, to put our results in a more recognisable form, we compute the coordinate transformation that sends observational to comoving coordinates. To do so requires specifying a gauge for $r$. We have chosen the gauge $r = uD$ on the PLC0 since it recovers the most well known form of the comoving FLRW metric viz. 
\begin{equation}
ds^2 = -dt^2 + a(t)^2\left(\frac{dr^2}{1 - Kr^2} + r^2d\Omega^2\right).
\label{FLRWMet}
\end{equation}
If the universe is truly FLRW then the comoving coordinates we transform to are exactly the same as those in the above metric. In particular note that for a flat FLRW universe this gauge implies $X(t,r) = a(t)$ and $R(t,r) = a(t)r$. Thus on any constant time slice the metric function in front of $dr^2$ should be constant and the one in front of $d\Omega^2$ should be a linear function of the comoving distance $r$. To further disentangle these two models we also identify functions that are a priori different in $\Lambda$LTB and FLRW universes.

\subsection{Discriminating between models}\label{sec:DisMod}
A number of consistency relations have been proposed to test the assumption of homogeneity \cite{Maartens:2011ITUH}. Here we review the fundamental differences between FLRW and $\Lambda$LTB models and show how they can be used to test for radial inhomogeneity. We will assume that $D(z)$ and $H_{\parallel}(z)$ as well as all their derivatives w.r.t $z$ are known on the PLC0. We further assume that the metric components of \eqref{ObsMet1} are known inside the PLC. The values of these quantities are all readily available once the CIVP has been solved (see section~\ref{sec:CIVP}). We proceed with a discussion about kinematics in spherically symmetric dust space-time. \\
First of all using $u_au^a = -1$ we write the four velocity as (recalling that $u^0 = (1+z) = u$)
\begin{equation}
u^a = [u,\frac{Au^2-1}{2u},0,0].
\label{FourVel}
\end{equation}
Next a first order Taylor expansion on $z(w_0 + \delta w, v_0 + \delta v)$ allows us, upon taking the limit of infinitesimal translations, to write
\begin{equation}
\frac{dz}{dw} = \frac{\partial z}{\partial w}|_{v = \mbox{const.}} + \frac{\partial z}{\partial v}|_{w = \mbox{const.}}\frac{dv}{dw}.
\label{RedDrift0}
\end{equation}
Recognising the partial derivatives $\frac{\partial z}{\partial w}|_{v = \mbox{const.}} = \dot{u} = \dot{u}_1, ~ \frac{\partial z}{\partial v}|_{w = \mbox{const.}} = u^2 H_\|$ and using the metric \eqref{ObsMet1} to express $\frac{dv}{dw}$ we arrive at the following expression for the redshift drift
\begin{equation}
\frac{dz}{dw} = \dot{u} + \frac{1}{2}u^2H_\|\left(A - \frac{1}{u^2}\right).
\label{RedDrift1}
\end{equation}
In section~\ref{sec:algorithm} we show how this fact can be used to infer the value of $\Lambda$. \\
Next we can decompose the ray 4-vectors into parts parallel and orthogonal to $u^a$ as
\begin{eqnarray}
k^a &=& (-u_bk^b) \left(u^a + e^a\right) = - u \left(u^a + e^a\right), \\
&\mbox{where}& \quad e_ae^a = 1 \quad \mbox{and} \quad e_au^a = 0. \nonumber
\end{eqnarray} 
Here $e^a$ is the direction of propagation of the ray as seen by $u^a$ i.e.
\begin{equation}
e^a = \frac{k^{<a>}}{(-u_bk^b)} = [-u, -\frac{1 + Au^2}{2u},0,0].
\end{equation}
Using this information to compute $H_{obs} = \frac{1}{3}\Theta + \sigma_{ab}e^ae^b$ gives (note that in $\Lambda$LTB models $H_{obs} = H_\| = \frac{\partial_t X}{X}$)
\begin{equation}
H_{obs} = \dot{u} + \frac{u'}{2u^2} + \frac{u'A + uA'}{2}.
\label{Hpar}
\end{equation}
Further, projecting the null geodesic equation along the direction of propagation of the ray gives an expression that relates $v$ to $z$ viz. 
\begin{equation}
e^ak^b\nabla_b k_a = 0, \quad \Rightarrow \quad \frac{dz}{dv} = u^2 H_{\|},
\end{equation} 
which can be used to get the affine parameter as a function of redshift 
\begin{equation}
v(z) = \int_0^z \frac{dz}{(1+z)^2H_\|(z)}.
\label{nuz}
\end{equation}
The expression for the expansion scalar $\Theta = \nabla_a u^a$ is
\begin{equation}
\Theta = \dot{u} + \frac{uA'}{2} + \frac{u'A}{2} + \frac{u'}{2u^2} + \frac{2u\dot{D}}{D} - \frac{D'}{uD} + \frac{uD'A}{D}. 
\end{equation}
The transverse expansion rate is found by noting that in spherical symmetry $\Theta = H_{\|} + 2H_{\perp}$. Thus we have that
\begin{equation}
H_{\perp} = \frac{1}{D}\left(u\dot{D} - \frac{D'}{2u} + \frac{uAD'}{2}\right). 
\end{equation}
The limiting behaviour for these expressions as $v \rightarrow 0$ can be used to deduce that $H_\perp(w,0) = H_\|(w,0)$ i.e. the expansion rates are equal along $\mathcal{C}$. This fact allows us to compute $H_\perp$ without resorting to series expansions. More importantly it allows to get the current age of the universe by evaluating \eqref{gett0} without direct observations of $H_\perp$. In the remainder of the paper we use $H$ to refer to the parallel expansion rate $H_\|$ and explicitly write $H_\perp$ for the transverse expansion rate. Note that the matter shear $\sigma$, which is identically zero in FLRW models, is related to the difference between $H$ and $H_\perp$. We can use this information to formulate a test for radial inhomogeneity as follows
\begin{equation}
T_1 = 1 - \frac{H_{\perp}}{H}.
\label{sheartest}
\end{equation}
From the way $T_1$ is defined it is clear that it is a dimensionless quantity that should be zero in FLRW models. If it was possible to disentangle $H$ from $H_\perp$ observationally we could formulate a very direct test based on this quantity (for a related test see \cite{Maartens:2011ITUH},\cite{Clarkson:2007pz}). \\
Also, as proposed in \cite{Clarkson:2007pz}, we can formulate a test of the CP that is in principle independent of matter content and the particular theory of gravity employed by investigating the consistency between distances and the expansion rate. The two main geometric effects on distance measurements are:
\begin{enumerate}
\item curvature bends null geodesics,
\item expansion changes radial distances.
\end{enumerate}
In FLRW these two effects are coupled via the relation (assuming distance duality $d_L = u^2D$, where $d_L$ is the luminosity distance)
\begin{equation}
D = \frac{1}{u H_0\sqrt{-\Omega_{K0}}} \sin\left(\sqrt{-\Omega_{K0}}\int_0^z dz^* \frac{H_0}{H(z^*)}\right).
\end{equation}
This gives $\Omega_{K0}$ in terms of $H(z)$ and $D(z)$ as
\begin{equation}
\Omega_{K0} = \frac{\left[H\left(u D_{,z} + D\right)\right]^2 - 1}{\left[H_0 u D\right]^2}. 
\end{equation}
Since $\Omega_{K0}$ is expected to be independent of $z$ this expression can be derived to yield a quantity that should be zero in FLRW models viz.
\begin{eqnarray}
T_2 = 1 &+& H^2\left[u^2(DD_{,zz} - D_{,z}^2) - D^2\right] \nonumber \\
  &+& uHH_{,z}D\left[uD_{,z} + D\right], \label{curvetest}
\end{eqnarray}
where we use $H = H_\|$ since it is the radial expansion that effects radial geodesics. Note that although this quantity is in principle independent of the matter content and theory of gravity employed, the quantity that we reconstruct is not. This is because we use one of the field equations to constrain $D(z)$ and its derivatives. Accurately correlating and reconstructing derivatives of $D(z)$ and $H(z)$ in a non-parametric way is a near impossible task otherwise. \\
In what follows we will be using the quantities $T_1$ and $T_2$ to measure the degree of radial inhomogeneity allowed by the data. Confirming that either of these quantities are zero, or at least small enough to be consistent with FLRW + perturbations, validates the Copernican principle against the data employed.

\subsection{Characteristic initial value problem}\label{sec:CIVP}
Here we simply give a superficial outline of the CIVP formalism and refer the reader to \cite{Bester:2013fya} and \cite{prd2012} for further details regarding its implementation. The field equations are
\begin{eqnarray}
D'' &=& -\frac{1}{2}\kappa D \rho (u_1)^2, \label{RNU} \\
\dot{D}' &=& \frac{1}{2D} \left[1- DD'A' - 2\dot{D}D' - A(D')^2 \right. \nonumber \\
&~& \quad \quad \left. - ADD'' - \frac{1}{2}\kappa \rho D^2 - \Lambda D^2\right], \label{RW} \\
A'' &=& \kappa A(u_1)^2 \rho - 4\frac{\dot{D}'}{D} -2\frac{A'D'}{D} - 2\Lambda,  \label{WNU}\\
D(0) &=&  A'(0) = \dot{D}(0) = 0, ~~ A(0) = D'(0)=1. \label{ICs}
\end{eqnarray}
These can be considered as constraint equations that need to be solved on each PLC. Equation~\eqref{RNU} in particular plays a very important role in setting up the data on the PLC0 since it ensures that all physical constraints are satisfied. Assuming that we have the functional forms of $H(z)$ and $\rho(z)$, as well as the value of $\Lambda$, the $\Lambda$LTB solution on the PLC0 can be found with \ref{AlgINIT}. 
\begin{algorithm}
\SetAlgoRefName{Procedure 1}
\NoCaptionOfAlgo
\nl $H(z)$ with \eqref{nuz} gives the $v(z)$ relation which gives $\rho(v)$ and since $v(z)$ is invertible also $u_1(v) = 1 + z(v)$\;
\nl Solve \eqref{RNU} with $u_1(v)$ and $\rho(v)$ to find $D(v)$\;
\nl Use \eqref{gett0} to get $t_0$ and write initial data to the grid on which the solution is desired\;
\nl Solve the coupled system \eqref{RW} and \eqref{WNU} for $A(v)$ and $\dot{D}(v)$ and use \eqref{U1W} to evaluate $\dot{u}_1$.
\caption{Procedure 1 \label{AlgINIT}}
\end{algorithm}
This is all that is required to compute the potential function $\Phi$ of \eqref{RNBayThe} and thus confront the current realisations of $H(z), \rho(z)$ and $\Lambda$ with data.\\
To find the solution on the inside of the PLC we need a way to evolve the initial data to the next PLC. To do so we need to specify, in addition to the output from \ref{AlgINIT}, the values of $u_0, \dot{u}_{1}$ and $\dot{\rho}$. The value of $u_0$ can be found from the normalisations condition $u_au^a = -1$ whereas $\dot{u}_{1}$ and $\dot{\rho}$ follow from the conservations equations $\nabla_bT^b_{\ a} = 0$:
\begin{eqnarray}
\dot{u}_{1} &=& \frac{1}{2}\left[(\frac{1}{(u_1)^2} - A)(u_{1})' -A'u_1\right], \label{U1W}\\
\dot{\rho} &=& -\frac{1}{u_1}\left[\rho \big( \dot{u}_{1} + (u_{0})' + 2A'u_1 + 2\frac{\dot{D}}{D}u_1 + A (u_{1})' \right. \nonumber \\
&+& \left. 2\frac{D'}{D}(u_0+A u_1)\big)  +\rho'(u_0 + A u_1) \right]. \label{RHOW}
\end{eqnarray}
There is still one final piece of information required to solve this system. As mentioned in section~\ref{sec:Model} the coordinate $v$ is necessarily non-comoving. This means that its maximum extent changes as we go from one PLC to the next, its maximum extent being determined by the causal horizon or characteristic cut off line (henceforth $\mathcal{W}$). In order to compute this characteristic cut off we compute the path of a null geodesic from the maximum value of $v$ into the interior of the PLC (see the discussion in \cite{Bester:2013fya} for further details about how this is implemented). In principle anything beyond $\mathcal{W}$ has never been in causal contact with the observable universe. In a numerical application however, since at least two grid points are required to take derivatives, there is always a tiny portion of the PLC that is beyond our reach i.e. the intersection of $\mathcal{W}$ with $\mathcal{C}$. This limitation is unavoidable but note that the grid can be chosen fine enough so that these two grid points lie well within the averaging scale and are therefore of no practical importance. The numerical error of the integration scheme has been set to $10^{-6}$ throughout. Since the scheme is second order the affine parameter distance between these two grid points is $v \approx \sqrt{10^{-6} ~ \mbox{Gpc}} = 1$ Mpc. \\
Finally, given the output from \ref{AlgINIT} above, the full $\Lambda$LTB solution in observational coordinates can be found using \ref{AlgCIVP}. To find the solution in comoving coordinates we need to specify the coordinate transformation.
\begin{algorithm}
\SetAlgoRefName{Procedure 2}
\NoCaptionOfAlgo
\nl Get $u_0$ from $u_au^a = -1$ and evolve the system to the next PLC by solving \eqref{U1W} and then \eqref{RHOW}\;
\nl Solve \eqref{RNU}, \eqref{RW} and \eqref{WNU} on this PLC\;
\nl Repeat until the domain of calculation is exhausted\;
\nl Get the characteristic cut off line $\mathcal{W}$.
\caption{Procedure 2 \label{AlgCIVP}}
\end{algorithm}

\subsection{Coordinate Transformation}
\begin{figure}
\includegraphics[width=0.45\textwidth]{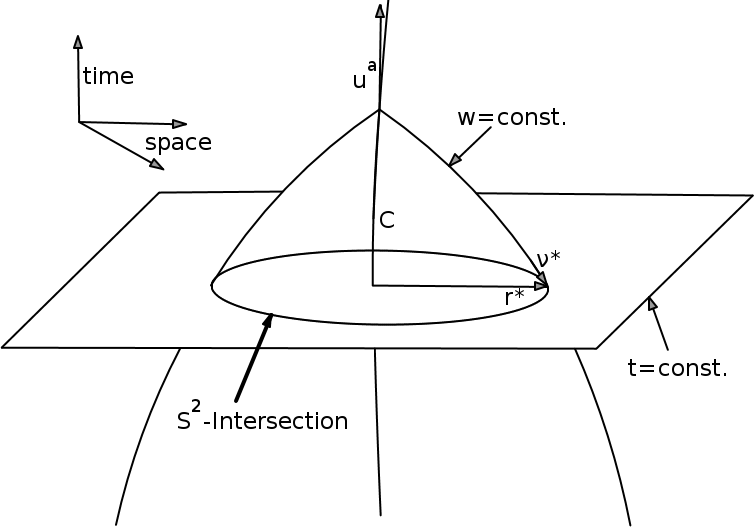}
\caption{An illustration of the intersection of the two coordinate systems. It is only possible to reconstruct constant time surfaces that lie within the 2 sphere of intersection of the coordinate systems. \label{fig:CoordInt}}
\end{figure}

Here we derive the transformation that links observational to comoving coordinates (schematically illustrated in Figure~\ref{fig:CoordInt}). If $t = t(r)$ is the solution for $t$ on the PLC0 then choosing $R(t(r),r) = D(w_0,\upsilon)$ ensures that $\theta$ and $\phi$ have the same meaning in both coordinate systems. Thus we only need to consider the $(w,\upsilon) \leftrightarrow (t,r)$ transformation. The CIVP formalism solves for the background cosmological metric \eqref{ObsMet1} in terms of observational coordinates $x^{a}$. We would like to compare these solutions to known solutions of the metric \eqref{LTBMet1} in terms of comoving coordinates $x^{\tilde{a}}$. Accordingly we need to find both the metric components of \eqref{LTBMet1} as functions of $x^a$ and then explicitly solve for comoving coordinates in terms of $x^a$. The metric components follow from the transformation law
\begin{equation}
g_{\tilde{a}\tilde{b}} = \frac{\partial x^{c}}{\partial x^{\tilde{a}}}\frac{\partial x^{d}}{\partial x^{\tilde{b}}}g_{cd}.
\label{ForTrans}
\end{equation}
Clearly we need expressions for these partial derivatives purely in terms of the observational metric and coordinates. Once we have the comoving metric in observational coordinates the geodesic equations can be solved on each PLC to explicitly find comoving coordinates in terms of observational coordinates.\\
Gauge fixing the coordinate $w$ to measure proper time along the central worldline means that the partial derivatives involving time are straightforward. Using that $dt = -u_adx^a$, $dw = k_adx^a$ and $u^1 = \upsilon_{,a}u^a = \frac{\partial \upsilon}{\partial t}$ gives (where we have used the expression for $u^1$ from \eqref{FourVel})
\begin{eqnarray}
\frac{\partial t}{\partial w} &=& \frac{Au^2 + 1}{2u}, \qquad \frac{\partial t}{\partial v} = - u,  \label{tder1} \\
\frac{\partial w}{\partial t} &=& u \qquad \mbox{and} \qquad  \frac{\partial v}{\partial t} = \frac{Au^2 - 1}{2u}. \label{tder2}
\end{eqnarray}
The partial derivatives involving $r$ require a little more effort. The transformation \eqref{ForTrans} as well as its inverse gives (where $a = \frac{\partial w}{\partial r}$, $b = \frac{\partial v}{\partial r}$, $c = \frac{\partial r}{\partial w}$ and $d = \frac{\partial r}{\partial v}$ for notational simplicity)
\begin{eqnarray}
d^2X^2 = g^2d^2(a\dot{D}+bD')^2 &=& u^2, \label{tr1} \\
bu - \frac{1}{2}auA - \frac{a}{2u} &=&0, \label{tr2} \\
cu +\frac{1}{2}duA - \frac{d}{2u} &=& 0, \label{tr3} \\
ac + bd  &=& 1, \label{tr4} 
\end{eqnarray}
where we have used $X = g(r)\partial_r R = g(r)\left(a \dot{D}  + b D' \right)$. Since there are five unknowns in four equations some additional information is required to solve this system. This is provided by the fact that the partial derivatives in these transformations commute. No new information can be derived from applying this criterion to $a$ and $b$, it simply recovers the momentum conservation equation \eqref{U1W}. Applied to $c$ and $d$ however we get a partial differential equation (PDE) for $d$ viz. $c' = \dot{d}$. This casts \eqref{tr3} into the following flux conservative form
\begin{equation}
\dot{d} = -\frac{\partial}{\partial v}\left(\frac{d}{2}\left(A - \frac{1}{u^2} \right) \right).
\label{dPDE}
\end{equation}
Since $A$ and $u$ are known on the grid all that is required to solve this PDE are initial conditions for $d$. These are provided by fixing the gauge freedom in $r$. It is clearest to proceed with an FLRW analogy. To that end recall that in FLRW universes the comoving distance is simply $r=\frac{uD}{a_c}$, where $a_c$ is a constant equal to the value of the scale factor on $\mathcal{C}$. Thus in FLRW the transformation is extremely simple, $d$ is always given by
\begin{equation}
\frac{\partial r}{\partial \upsilon} = d = \frac{u'D + u D'}{a_c}.
\label{d0}
\end{equation}
Having fixed $d$ we may use \eqref{tr1}-\eqref{tr4} to find the remaining unknowns viz. $a, b, c$ and $X$. To find the transformation in general we should see \eqref{d0} as a normalisation on the PLC0. The initial data for $d$ then makes it possible to solve \eqref{dPDE} and the remaining unknowns again follow from \eqref{tr1}-\eqref{tr4}. Partial derivatives w.r.t. either $r$ or $t$ can now be expressed as derivatives w.r.t. $w$ and $v$ using straightforward tensor calculus. To get $t(w,v)$ and $r(w,v)$ we further need to solve the following geodesic equations on each PLC
\begin{eqnarray}
\frac{d^2 t}{d\upsilon^2} &=& -\frac{\partial_t X}{X} \frac{d t}{d\upsilon}^2  = - H\frac{d t}{d\upsilon}^2, \label{geot} \\
\frac{d^2r}{dv^2} &=& -\frac{\partial_r X}{X}\left(\frac{dr}{dv}\right)^2 - 2H \frac{dr}{dv}\frac{dt}{dv}. \label{geor}
\end{eqnarray}
These solutions allow us to associate the corresponding $(t,r)$ pair to any $(w,v)$ grid point. In what follows we are going to reconstruct a snapshot of the universe at a fixed time, $t = t^*$ say, in the past. Let us denote the hypersurface defined by $t^* = $ const. by $\Sigma_t$ and the radial coordinate on this hypersurface by $r^*$. Then our aim is to find the intersection of functions defined on the $(w,v)$ grid with $\Sigma_t$. This can be achieved by finding the intersection of each PLC for which $w \geq t^*$ with $\Sigma_t$ and using the coordinate transformation to associate these points of intersection, $(w^*,v^*)$ say, to the corresponding comoving coordinates. This is implemented using \ref{AlgTransform}. 
\begin{algorithm}
\SetAlgoRefName{Procedure 3}
\NoCaptionOfAlgo
\nl Use \eqref{tder1} and \eqref{tder2} to get partial derivatives w.r.t $t$ \;
\nl Solve \eqref{dPDE} for $d$ with initial data set by \eqref{d0} and use \eqref{tr1}-\eqref{tr4} to find $a,b,c$ and $X$\;
\nl Solve \eqref{geot} and \eqref{geor} on each PLC to get $t(w,v)$ and $r(w,v)$ on the grid.\;
\nl By construction $w^*$ consists of all the grid points for which $w \geq t^*$. Thus for each $w^*$ find $v^*$ such that $t(v^*) = t^*$ and then interpolate $r(v)$ to find $r^* = r(v^*)$\;
\nl Having located $(w^*,v^*)$ simply interpolate the desired functions and report their values at the corresponding $(t^*,r^*)$.
\caption{Procedure 3 \label{AlgTransform}}
\end{algorithm}

\section{Smoothing the data}\label{sec:SmoothData}
\subsection{Gaussian process regression}\label{sec:GPR}
Gaussian processes have recently become quite popular as a means to perform non-parametric regression on cosmological data sets (see for example \cite{Seikel:2012uu,2012PhRvD..85l3530S,2014IAUS..306...25B,2014MNRAS.441L..11B}). As a result we will only give a superficial outline of the theory relevant to the current application. In its most basic form a GP is a collection of random variables. Any finite collection of these variables have a joint Gaussian distribution \cite{rasmussen2006gaussian}. A Gaussian process can be completely characterised by specifying its mean $m(\mathbf{x})$ and covariance function $k(\mathbf{x},\tilde{\mathbf{x}})$. The mean and covariance function of a real process $f(\mathbf{x})$ are defined by
\begin{eqnarray}
m(\mathbf{x}) &=& \mathbb{E}[f(\mathbf{x})], \\
k(\mathbf{x},\tilde{\mathbf{x}}) &=& \mathbb{E}[(f(\mathbf{x}) - m(\mathbf{x}))(f(\tilde{\mathbf{x}}) - m(\tilde{\mathbf{x}}))], 
\end{eqnarray}
where $\mathbb{E}[\mathbf{x}]$ denotes the expectation value of $\mathbf{x}$ with respect to a Gaussian distribution. This is conveniently abbreviated using the notation $f(\mathbf{x}) \sim \mathcal{G}\mathcal{P}(m(\mathbf{x}),k(\mathbf{x},\tilde{\mathbf{x}}))$. We will further abbreviate our notation by labelling the Gaussian processes for the different functions using subscripts. Thus $\mathcal{GP}_H$ and $\mathcal{GP}_\rho$ refer to the Gaussian processes for the expansion rate and density respectively.\\
The Gaussian process regression (GPR) problem consists of making inferences about the relationship between the inputs and the targets. The covariance matrix follows from evaluating the covariance function at the relevant points i.e. $K_{ij} = k(x_i,x_j)$. Denoting the prior distribution over functions as $f_p$, the joint distribution between $f_p$ and the data is given by
\begin{equation}
\left[ \begin{array}{c} y \\ f_p \end{array}\right] \sim  \mathcal{N}\left( \mathbf{0}, \left[ \begin{array}{cc}
K(X,X) + \Sigma & K(X,X_p) \\ K(X_p,X) & K(X_p,X_p) \end{array}  \right] \right),
\end{equation} 
where $\Sigma$ is the covariance matrix the of the data, $X$ is the vector of inputs and $X_p$ the vector of targets. The posterior distribution is the distribution of $f_p$ restricted to be compatible with the observations or, in probabilistic terms, the conditional distribution of $f_p$ given data and targets i.e. 
\begin{eqnarray}
f_p&|&X,\mathbf{y},X_p \sim \mathcal{N}\left(\bar{f}_p, \cov(f_p)\right), ~~ \mbox{where} \label{pred1} \\
\bar{f}_p &=& K(X_p,X)K_y^{-1} \left(y - m(X)\right), \label{predmean} \\
\cov(f_p) &=& K(X_p,X_p) - K(X_p,X)K_y^{-1}K(X,X_p). ~~ \label{predvar}
\end{eqnarray}
Here $\bar{f}_p = \mathbb{E}[f_p|X,y,X_p]$ is the posterior mean, $\cov(f_p)$ the posterior covariance matrix and $K_y = [K(X,X) + \Sigma]$. The marginal log-likelihood associated with GPR is given by 
\begin{equation}
\log(p(y|X,\theta)) = -\frac{1}{2}y^T K_y^{-1} y - \frac{1}{2} \log|K_y| - \frac{n}{2}\log(2\pi), 
\label{MarPos}
\end{equation}
where $\theta$ is a vector of hyperparameters. Eqns \eqref{pred1} - \eqref{predvar} are the key predictive equations for GPR.\\
We employ the Mattern class of covariance functions with $\nu = 5/2$ throughout (note $r = |x-\tilde{x}|$)
\begin{equation}
k(x,\tilde{x}) = \left(1+\frac{\sqrt{5}r}{l} + \frac{5r^2}{3l^2}\right) \sigma_f^2 \exp\left(-\frac{\sqrt{5}r}{l}\right).
\label{Mat52}
\end{equation}
Mean functions are chosen using an iterative procedure that will be discussed in section~\ref{sec:Priors}.

\subsection{Enforcing physical constraints}\label{sec:physcons}
Substituting $\frac{dz}{dv} = u^2 H$ into the chain rule applied to $D'$ shows that the angular diameter distance and the expansion rate are related by
\begin{eqnarray}
D' &=& \frac{dz}{dv}\frac{dD}{dz} =  u^2 H D_{,z}, \label{Dp} \\
D'' &=& u^3\left( 2H^2 D_{,z} + uHH_{,z}D_{,z} + uH^2 D_{,zz}\right). \label{Dpp}
\end{eqnarray} 
Our reasons for showing these equations are twofold. Firstly they show that incorporating the interdependence between $D(z)$ and $H(z)$ while smoothing them separately is non-trivial. It is far easier to start with realisations of $H(z)$ and $\rho(z)$ and then to find $D(z)$ as described in \ref{AlgCIVP}. We then infer $D(z)$ from the available data as explained in \ref{Alg1} below. Note that regularity at the vertex of the cone is enforced by solving \eqref{RNU} with the specified initial conditions \eqref{ICs}. Secondly these expressions can be used to minimise the number of additional numerical derivatives that need to be found in order to reconstruct $T_2$ of \eqref{curvetest}. In fact we only need to numerically differentiate $H$ w.r.t. $z$, the other required derivatives (towards $z$) then follow from \eqref{Dp} and \eqref{Dpp}. \\
The null energy condition can be enforced simply by rejecting samples with $\rho < 0$. To avoid shell crossing singularities we also need to ensure that the density remains regular as $\partial_r R(t,r)$ crosses zero \cite{PhysRevD.90.064021}. This can only be done post integration and once we have the form of the coordinate transformation. We explicitly check this condition and discard the sample if it does. However since we sample $\rho(z)$ directly we found that this virtually never happens.

\section{The algorithm}\label{sec:algorithm}
Let us again emphasise that a spherically symmetric dust universe has two free functions. In our algorithm we select $\rho(z)$ and $H(z)$ as the two free functions. Given a value of $\Lambda$ we then solve the system of field equations by implementing \ref{AlgINIT} followed by \ref{AlgCIVP}. Since the entire solution is known numerically on a grid, and we can convert any function to a function of redshift, we are able to assign a likelihood to the current sample of $H(z)$, $\rho(z)$ and $\Lambda$ by confronting this solution with the available data. Importantly we can use any of the available data sets to compute this likelihood. However we should point out that using only a single data set will tend to favour models of the universe that only have a single free function. It is therefore imperative that we use at least two data sets to perform inference. Further, as evident from \eqref{RNU}, density data can be used to further constrain $D(z)$ and $H(z)$ but it cannot tell us anything about $\Lambda$ since $\Lambda$ has not yet entered the picture. Also, from \eqref{RW}, \eqref{WNU} and \eqref{U1W}, the computation that leads to the form of $\dot{u}_1(z)$ involves all of $D(z)$, $H(z)$, $\rho(z)$ and $\Lambda$. Redshift drift data will therefore have the most constraining power and is required to ultimately test the Copernican principle. In full awareness of this limitation the best we can currently do is to marginalise over the value of $\Lambda$ by sampling it from some suitable prior. As already mentioned prior specification is an important aspect of the algorithm so we will discuss that before presenting the algorithm used to perform inference.  
\begin{figure*}
\includegraphics[width=1.0\textwidth]{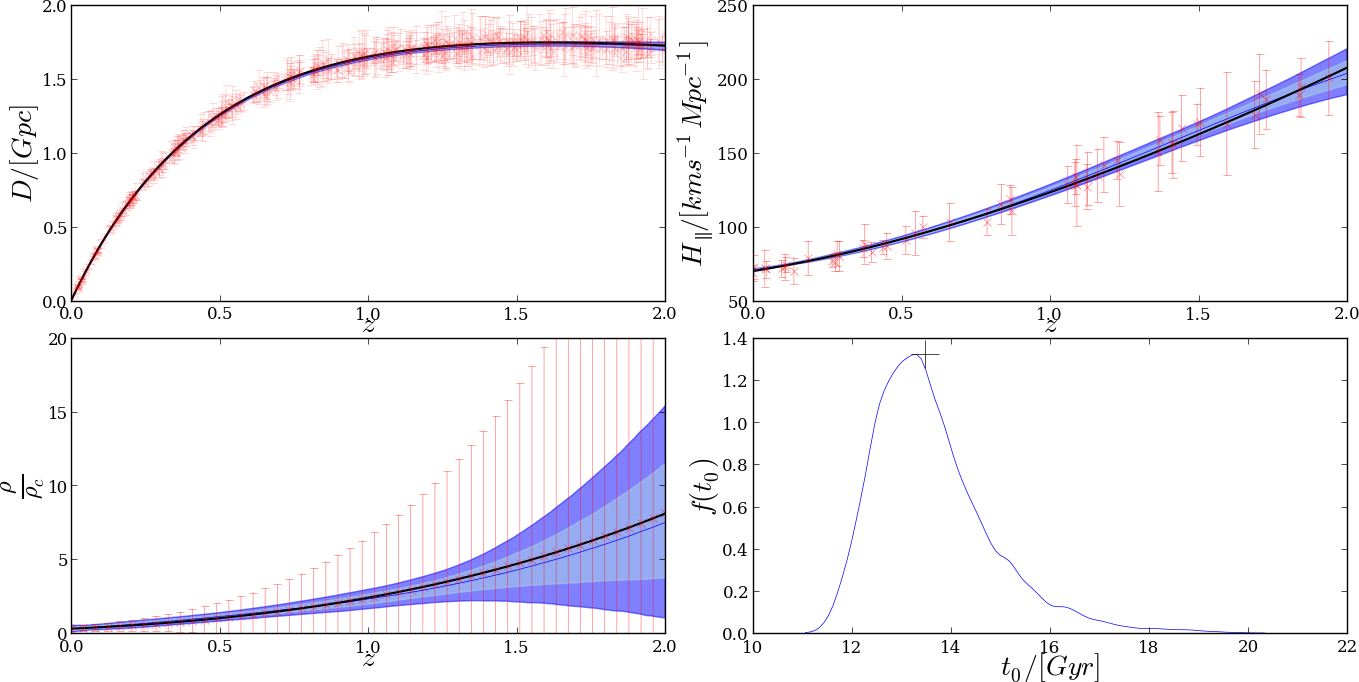}
\caption{Observables on the PLC0 from simulated data. Top Left: Posterior distribution of $D(z)$ on the PLC0. The contours are of similar width as when smoothing the $D(z)$ alone except at the far end of the data where they are drastically reduced. This is because concave realisations of $D(z)$ are excluded on the basis that they lead to violations of the null energy condition. Top Right: Posterior distribution of $H(z)$ on the PLC0. The contours are significantly narrower than when smoothing $H(z)$ data alone, especially for small $z$. The reason for this is that, as can clearly be seen from \eqref{Dp} and \eqref{Dpp}, the $D(z)$ data are able to significantly constrain $H(z)$. Bottom Left: The posterior distribution of $\rho(z)$ on the PLC0. Clearly the error bars far overestimate the posterior uncertainty in the reconstructed $\rho(z)$. Bottom Right: The posterior distribution of $t_0$. Note how close the peak is to the background $\Lambda$CDM value indicated by the black cross in the figure. Also $t_0$ is quite well constrained, virtually none of the models have $t_0 < t^*$. \label{fig:SimPLC0}}
\end{figure*}
\subsection{Prior specification}\label{sec:Priors}
Depending on the data available we might wish to specify priors in a number of different ways. The efficiency of our algorithm is vastly improved by knowing approximately which functions to look for. Specifically we use data to inform our priors when such data are available. In the absence of data we use our best guess and fine tune it by looking at the diagnostics of the MCMC. Of course there are many other ways in which priors could be specified, there is no claim that our choices are optimal in any sense. Actually a number of improvements, eg. \emph{sieve priors} \cite{cotter2013}, are possible. These will be investigated in future research. \\
Note that the priors we specify below will also be used as proposals for the MCMC described in section~\ref{sec:inference}. 
\begin{figure*}
\includegraphics[width=1.0\textwidth]{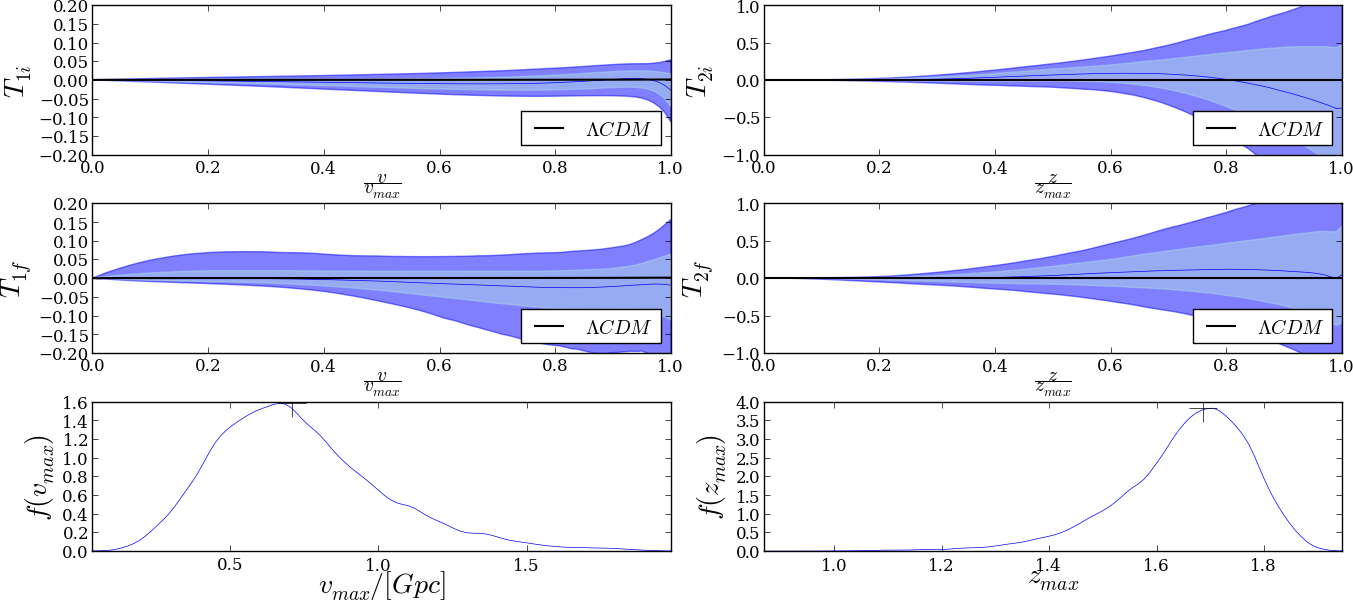}
\caption{The quantities $T_1$ and $T_2$ from simulated data. Top left: Posterior distribution of $T_1$ on the PLC0 marginalised over $t_0$. In order to compare successive realisations we plot $T_1$ as a function of the normalised affine parameter i.e. $\frac{v}{v_{max}}$. Middle Left: Posterior distribution of $T_1$ on the PLCF similarly plotted against the normalised affine parameter. Bottom Left: The maximum extent of $v$ on the PLCF. Note how close the peak falls to the background $\Lambda$CDM value indicated by the black cross in the figure. Top Right: Posterior distribution of $T_2$ on the PLC0 marginalised over $t_0$. In order to compare successive realisations we plot $T_2$ as a function of the normalised redshift i.e. $\frac{z}{z_{max}}$. Middle Right: Posterior distribution of $T_2$ on the PLCF plotted against the normalised redshift. Bottom Right: The maximum extent of $z$ on the PLCF. Again the peak is very close to the background $\Lambda$CDM value indicated by the black cross in the figure. 
\label{fig:SimTCP}}
\end{figure*}
\subsubsection{Sampling $\Lambda$}
Prior specification for the parameter $\Lambda$ should be simple when redshift drift data become available. Since $\Lambda$ is only a parameter any 1D distribution should suffice. It would be simplest to incorporate a Gaussian prior into our algorithm because then we could simply augment the vector $x$ in \eqref{vecx} with $\Lambda$ and use the MCMC diagnostics to pin down its mean value and fine tune its variance. \\
Since we cannot infer the value of $\Lambda$ without redshift drift data we treat it as nuisance parameter. Depending on the application we use two separate priors. Firstly we do not want the uncertainty in $\Lambda$ to obscure our results on simulated data. We therefore choose an accurate Gaussian prior when our primary intention is to verify that the algorithm works as expected on simulated data. In particular we use the background $\Lambda$CDM value with 2\% uncertainty. \\
When working with real data however we want to derive constraints regardless of the value of $\Lambda$. To do this we sample $\Lambda$ from a uniform distribution $\Lambda \sim \mathcal{U}[0,\Lambda_{max}]$ where $\Lambda_{max} = 3 (H^+_0)^2$. Here $(H^+_0)$ is the 2-$\sigma$ upper bound on the posterior of the expansion rate at the origin. Thus $\Lambda_{max}$ is the 2-$\sigma$ upper bound on $\Lambda$ when $\Omega_{\Lambda} = 1$.     

\subsubsection{Sampling $H(z)$}\label{sec:SamH}
There are a number of things to consider when setting a prior over functions for which data are available, two of which stand out in particular. Firstly, in order to avoid model misspecification as much as possible, we do not want to presuppose a parametrisation that might bias the space of functions considered. Secondly, it is important to explore the widest possible space of functions compatible with the data. With this in mind we used the following iterative procedure to set the prior on $H(z)$.\\
We start by performing GPR on expansion rate data using a zero mean function. We then use GaPP \cite{Seikel:2012uu} to train the hyperparameters of the Gaussian process. The optimised hyperparameters together with \eqref{predmean} give the posterior mean function. We then redo the GPR using this as the prior mean function and iterate until there is no significant change in the prior and the posterior mean functions. The posterior mean function for $H(z)$ resulting from the final iteration is denoted by $\bar{H}$. \\
In the Bayesian sense the widest space of functions compatible with the data actually results from marginalising over the hyperparameters. However we found that, especially for small data sets, better results were obtained on simulated data by using the optimised values resulting from the final iteration above. These values are therefore used to compute the mean function and covariance matrix with \eqref{predmean} and \eqref{predvar} respectively. We then use \eqref{pred1} to draw smooth function realisations and use these as input to the CIVP.

\subsubsection{Sampling $\rho(z)$} \label{sec:SamDen}
\begin{figure*}
\includegraphics[width=1.0\textwidth]{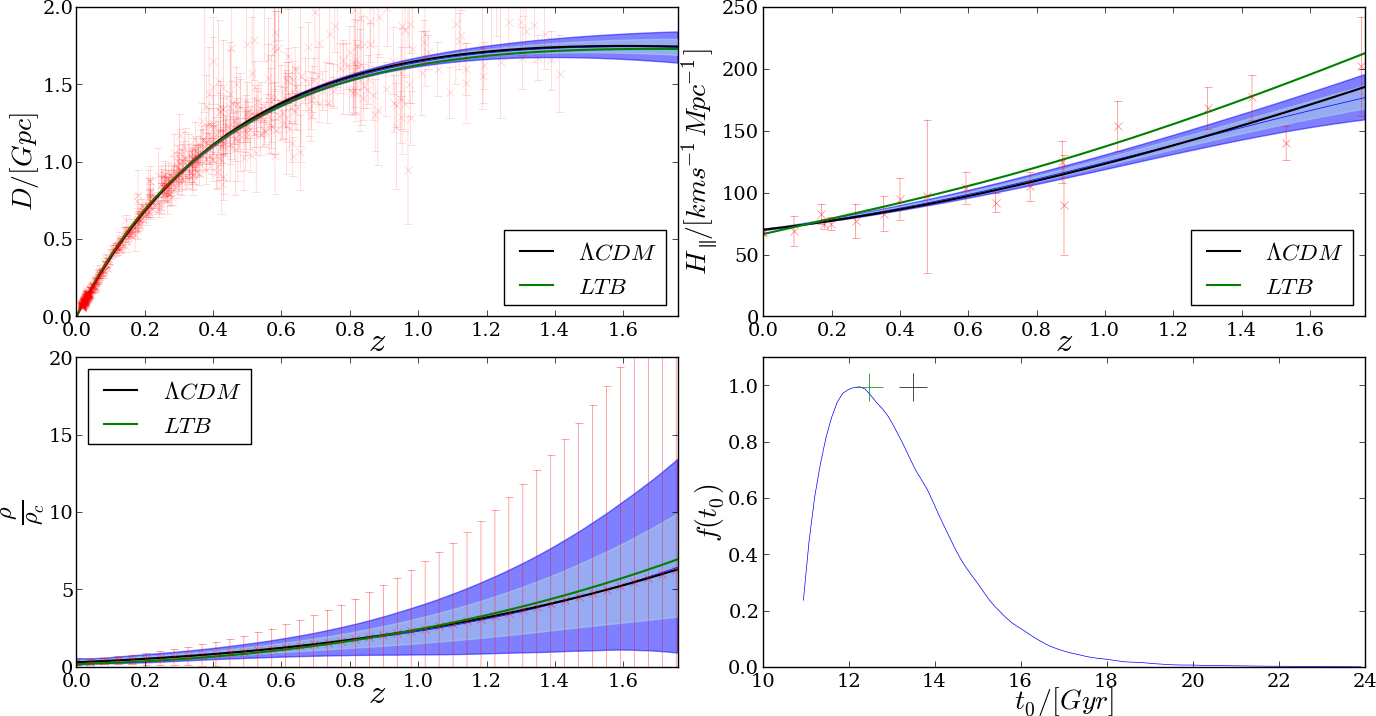}
\caption{Observables on the PLC0. Top Left: Posterior distribution of $D(z)$ on the PLC0 for real data out to $z = 1.75$. Top Right: Posterior distribution of $H(z)$ on the PLC0 for real data. Bottom Left: The posterior distribution of $\rho(z)$ on the PLC0 for real data. Bottom Right: The posterior distribution of $t_0$. The black (green) cross indicates the background $\Lambda$CDM (LTB) value. The large degree of uncertainty in $t_0$ mainly stems from the uncertainty in $\Lambda$. This results in a small percentage of the initial samples having a value of $t_0 < t^*$. For comparison we show, in green, these relations for a typical LTB model.  \label{fig:RealPLC0}}
\end{figure*}

Given density data, the procedure outlined above to sample $H(z)$ could simply be adapted to set a prior over $\rho(z)$. In the absence of data we have to be explicit about how the prior for $\rho(z)$ is set. Firstly we need to specify a mean function. For this we have implemented another iterative procedure in which we initially set the mean function as the background $\Lambda$CDM value. We then go through a number of burnin periods each time using the posterior median of $\rho(z)$ as the prior mean function in the next. This is repeated until there is no significant change in posterior median of $\rho(z)$. We have verified that specifying the initial mean function differently, as that corresponding to a void LTB model for instance, does not lead to a significantly different posterior median for $\rho(z)$. The posterior median of $\rho(z)$ resulting from the final iteration is denoted $\bar{\rho}$.\\  
Of course to implement this iterative procedure we need to model the prior covariance matrix of $\rho(z)$. For this we use a Gaussian process prior and then condition on artificial observations $z_i,\bar{\rho},\delta\rho_i$, where $\bar{\rho}$ gets updated after every burnin period. Within the GPR framework such a covariance matrix is given by
\begin{equation}
\Sigma_p = K_{pp} + K_p (K + \Sigma)^{-1} K_p^T,
\end{equation}
where $\Sigma$ is a matrix with variances along the diagonal. We have chosen to model these artificial observations so that their variance increases as a power law of the redshift i.e. $\delta \rho^2_i = \sigma^2_f(1+z_i)^\alpha$. Here $\alpha$ is a real number chosen in a very `liberal' manner so that the $\delta \rho_i$ completely overshoot the expected variance of $\rho(z)$. It is clear from Figure~\ref{fig:RealPLC0} that this is indeed the case.\\
The hyperparameters for $\mathcal{GP}_\rho$ are chosen in a somewhat ad hoc manner. For the length scale $l$ we initially pick a value that leads to reasonable density profiles. The signal variance $\sigma_f$ is then set to some small value and increased until we see that the prior distribution of $\rho(z)$ sufficiently covers its posterior. Both the length scale and the signal variance are adjusted iteratively to control the acceptance rate of the functional MCMC detailed in \ref{Alg1}. \\
This procedure might appear obscure and ad-hoc. However we found our results to be quite robust against this prior. Slightly altering the values of $l$ and $\sigma_f$ can effect on the overall acceptance rate of the MCMC but it does significantly change the final distributions of quantities $T_1$ and $T_2$ used to test the Copernican principle, only the time it takes for these distributions to converge.

\subsection{Inference}\label{sec:inference}
Here we describe the method used to perform inference. Angular diameter distance and expansion rate data only allow us to perform inference on $H(z)$ and $\rho(z)$ not on $\Lambda$. As such we construct the target vector
\begin{equation}
x = \left[ \begin{array}{c} H - \bar{H} \\ \rho - \bar{\rho} \end{array} \right],
\label{vecx}
\end{equation}
where $\bar{H}$ is the posterior mean function of $\mathcal{GP}_H$ and $\bar{\rho}$ the posterior median function of $\mathcal{GP}_\rho$. The quantity we want to infer from the available data is the posterior of $x$. To do so we implement a modified random walk method over the function space of $x$. A detailed discussion of MCMC algorithms for functional spaces is given in \cite{cotter2013}. We will restrict the current discussion only to the relevant theory. Note that $\mu(x)$ is used to refer to the measure on $x$ while $\mu_0(x)$ is used to refer to the measure on the prior over $x$.\\
One of the key difficulties with MCMCs over function spaces is that their acceptance rates are sensitive to spatial discritization. As the grid is refined acceptance rates typically drop. This is obviously far from ideal for numerical solutions to differential equations. The key idea to overcome this difficulty is to use, as proposals for Metropolis-Hastings type methods, discritizations of certain differential equations which exactly preserve the Gaussian reference measure as the likelihood drops to zero. In other words, in absence of evidence to the contrary, the posterior will be identical to the prior. Note that the prior $\mu_0(x)$ is now the joint normal distribution of the priors over $H$ and $\rho$. We assume that $H$ and $\rho$ are independent in the prior so that the off diagonal block matrices in the joint covariance matrix are just zero matrices. This effectively means that we can get a prior sample of $x$ by sampling $H(z)$ and $\rho(z)$ separately.\\
Next we note that if either the prior $\mu_0(x)$ or the posterior $\mu(x)$ can be chosen to be a dominating reference Gaussian measure, then Bayes' formula is expressed with the corresponding Radon-Nikodym derivative as
\begin{equation}
\frac{d\mu}{d\mu_0}(x) \propto L(x), \quad \mbox{where} \quad L(x) = \exp(-\Phi(x)).
\label{RNBayThe}
\end{equation} 
Here $L$ is the likelihood with corresponding real valued potential $\Phi$ chosen so that the Markov chain converges to the desired target. The current application dictates that we choose the prior to be the dominating reference Gaussian measure.\\ 
The preconditioned Crank-Nicolson (pCN) proposal of \cite{cotter2013} takes the form
\begin{equation}
y_{(k)} = \sqrt{(1-\beta^2)}x_{(k)} + \beta \delta, \quad \mbox{with} \quad \delta \sim \mu_0(x),
\label{pCNPro}
\end{equation}
where $0 \leq \beta \leq 1$ is a constant that can be adjusted to control the acceptance rate and we indicate the step in the chain by a subscript in braces. The acceptance probability is then defined as
\begin{equation}
a(x,y) = \min\left(1,\exp\left(\Phi(x) - \Phi(y)\right)\right).
\label{AccPro}
\end{equation}
This method differs from the standard random walk method since the proposal is not centred but rather autoregressive with order one ($AR(1)$ type). Importantly the method does not reject the proposal in the case where $\Phi = 0$ but rather accepts the move with probability one (which is why prior specification is so important). This is the key idea, the method exactly preserves the Gaussian reference measure because the proposal is reversible w.r.t. $\mu_0$. If we want to ensure that the MCMC rejects a sample, for example when certain physical constraints have been violated, then conceptually we should set $\Phi = \infty$. \\
Next we need to specify the potential function $\Phi$. For simplicity we use the joint chi-square of all the available data for this purpose 
\begin{equation}
\Phi(x) = \sum_i \chi^2_i.
\label{PotFunc} 
\end{equation}
Here $\chi^2_i$ is the chi-square value of data set $i$. For example if we have $j$ observations of the function $y$ then we use
\begin{equation} 
\chi^2_y = \frac{1}{2} \sum_j \frac{(y_j - \bar{y})^2}{ \sigma^2_{y_j}}. 
\end{equation}
\begin{figure*}
\includegraphics[width=1.0\textwidth]{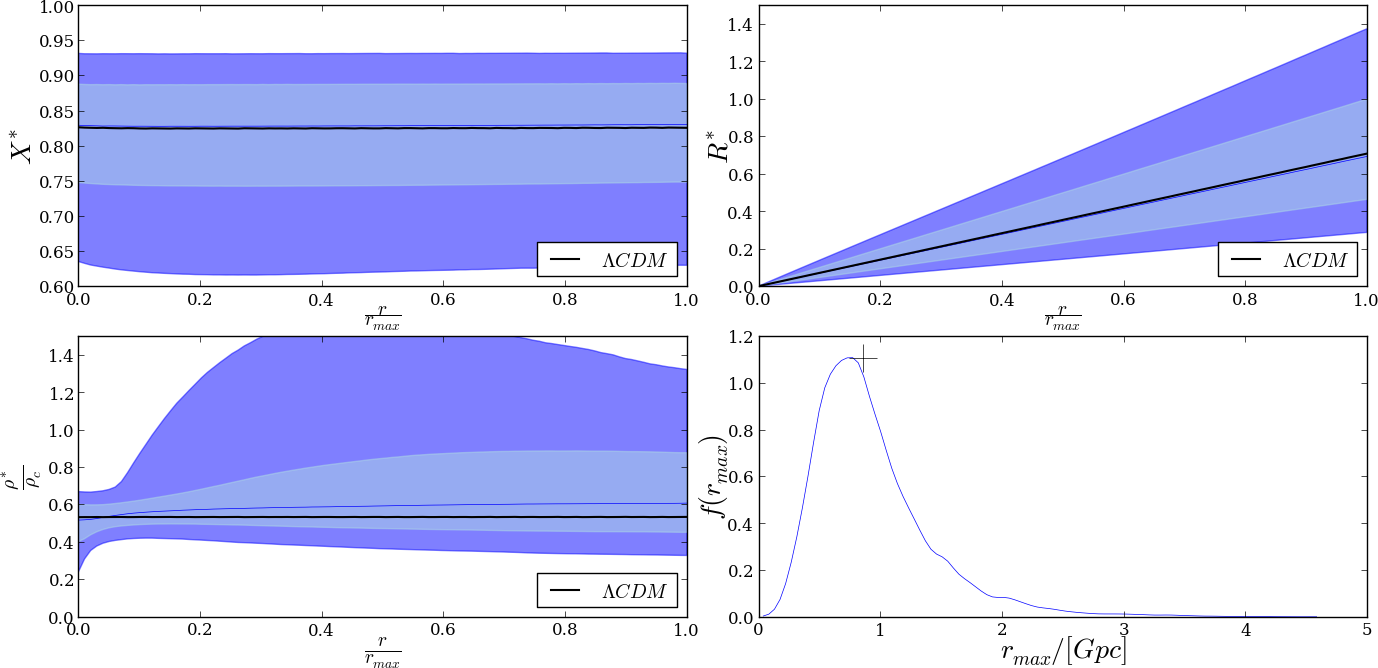}
\caption{Posterior distributions on a constant time slice $\Sigma_t$ defined by $t^* = 10.5$ Gyr for simulated data. We plot $X, R$ and $\rho$ as functions of the normalised comoving distance $\frac{r}{r_{max}}$. Top Left: As expected the metric function $X$ is very nearly constant. Bottom Left: The density on $\Sigma_t$ plotted in units of the critical density $\rho_c$ of the background $\Lambda$CDM model used in the simulations. Clearly the most probable density profile is that of a homogeneous matter distribution. Interestingly it seems that void profiles are not ruled out by the data. Top Right: Again as expected the metric function $R$ is very nearly a linear function of the comoving distance $r$. Bottom Right: The maximum extent of the radial coordinate $r^*$ on $\Sigma_t$. The peak of this distribution is very close to the background $\Lambda$CDM value indicated by the black cross in the figure.
\label{fig:Simtslice}}
\end{figure*} 
We could use a potential function other than the chi-square if we wanted to. Actually there is considerable flexibility in the way $\Phi$ can be specified and this is important for the intended application of the algorithm. As pointed out in the introduction it is substantially more difficult to obtain data that are valid in inhomogeneous cosmologies. This limits the applicability of our algorithm since there are very few model independent data available. For example there is an inherent circularity when incorporating BAO \cite{2010deot.book..246B}, weak gravitational lensing \cite{Bartelmann:1999yn}, redshift-space distortion \cite{redspacedistotions2011}, galaxy number count \cite{Gardner:1998ae} and galaxy cluster \cite{2011ARA&A..49..409A} data since they often assume a background FLRW cosmology at some point or another. Moreover, because of the complexity behind many of these astrophysical modelling techniques, the exact point at which such assumptions enter the calculation can be obscured to non-experts in any particular field. However, if we are able to pinpoint exactly where these assumptions enter the picture, it might still be possible to incorporate these data sets in a meaningful way. This would be possible if we can identify and reverse the effect that these assumptions have on the potential function $\Phi$. A similar idea is used to analyse CMB data in \cite{2010JCAP...08..023V} and \cite{Audren11102014}. Probably the simplest conceivable example would be to marginalise over the particular choice of $H_0$ when converting the Union 2.1 distance modulus to angular diameter distance data. However having admitted that current data are not able to place tight constraints on violations of the Copernican principle, we chose not to perform the marginalisation since that would only degrade the constraints further. Also note that $H_0$ is not the only nuisance parameter involved in analysing supernovae data \cite{2010ApJ...716..712A}. \\
It is unlikely that future surveys such as Euclid \cite{2013LRR....16....6A}, DES \cite{1742-6596-259-1-012080} or the SKA \cite{Maartens:2015mra} will produce data which are \emph{completely} model independent. We might therefore have to content ourselves with a certain degree of model dependence in the data. Realistic tests of the Copernican principle will therefore have to take special care to properly model the potential function $\Phi$ employed to perform inference. Another possibility is to use the model dependent data sets only as a means to specify
priors on certain functions and then use the model independent data to perform inference. These considerations will be addressed in future research.  

\subsection{Implementation}
In order to write down the final form of the algorithm we need to clearly state our objectives. The primary aim of this work is to test the Copernican principle. In order to do so we need to find the full distribution of solutions that are compatible with currently available data and then use these to reconstruct the quantities $T_1$ and $T_2$ defined in \ref{sec:DisMod}. However we still need to decide in advance on the domain over which the solutions should be found. Since the domain in $v$ is completely determined by the observations this amounts to specifying a time in the past up to which to integrate to. However this choice is not straightforward for two reasons. Firstly, the time is not arbitrary since, as explained in section~\ref{sec:CIVP}, $\mathcal{W}$ will intersect $\mathcal{C}$ at a finite time in the past. The available data therefore also places a limit on how far back in time we are able to go. Secondly, since there is also significant variation in the current age of the universe $t_0$, we do not simply want to choose a ``safe" value that will always lie within the causal horizon. The reason for this is that it might happen that a large percentage of the models considered end up having a $t_0$ smaller than the value we choose to integrate to. We therefore need to find a compromise between these two extremes. For this paper we have chosen to integrate up to $t^* \approx 10.5$~Gyr. Henceforth we refer to the PLC defined by $w = t^*$ as PLCF. We will also be reconstructing the distributions of $X, R$ and $\rho$ on the hypersurface $\Sigma_t$ defined by $t^* = $ const.\\
Having specified the domain we can find the full distribution of solutions inside the PLC up to the PLCF using \ref{Alg1}. For this algorithm we assume that initial realisations of $H(z), \rho(z)$ and $\Lambda$ have been chosen such that they satisfy all the physical requirements of section~\ref{sec:physcons}. The potential function is then initialised by computing \eqref{PotFunc} with the resulting model. \\
\begin{algorithm}
\SetAlgoRefName{Algorithm 1}
\NoCaptionOfAlgo
\nl Propose $y_{(k)} = \sqrt{1-\beta^2}x_{(k)} + \beta\delta, ~~ \delta \sim \mu_0(x)$ and sample $\Lambda$ \label{SamStep}\;
\nl \lIf{any($\rho < 0$)}{$\Phi = \infty$ go to \ref{AccStep}}
    \lElse{Implement \ref{AlgINIT}} 
	\Indp
\nl 	\lIf{ $t_0 < t^*$}{$\Phi = \infty$ go to \ref{AccStep}}
		\lElse{Implement \ref{AlgCIVP}}
		\Indp
\nl 		\lIf{$\mathcal{W}$ intersects $\mathcal{C}$ at $w > t^*$}{$\Phi = \infty$ go to \ref{AccStep}}
			\lElse{Implement \ref{AlgTransform}}	
			\Indp
\nl 			\lIf{Shell crossing}{$\Phi = \infty$ go to \ref{AccStep}}
				\lElse{Compute $\Phi(y_{(k)})$ with \eqref{PotFunc}}	
				\Indm \Indm \Indm
\nl Set $x_{(k+1)} = y_{(k)}$ with probability \eqref{AccPro} \label{AccStep}\;
\nl Set $k \leftarrow k+1$ and go to \ref{SamStep}\;
\caption{Algorithm 1 \label{Alg1}}
\end{algorithm}
The code implementing \ref{Alg1} is available on request and will be made publicly available at a later stage. It has been written in a mixture of Python and Fortran. This code was run on the Rhodes maths cluster which is hereby dubbed the \emph{GetaFix cluster}.

\begin{figure*}
\includegraphics[width=1.0\textwidth]{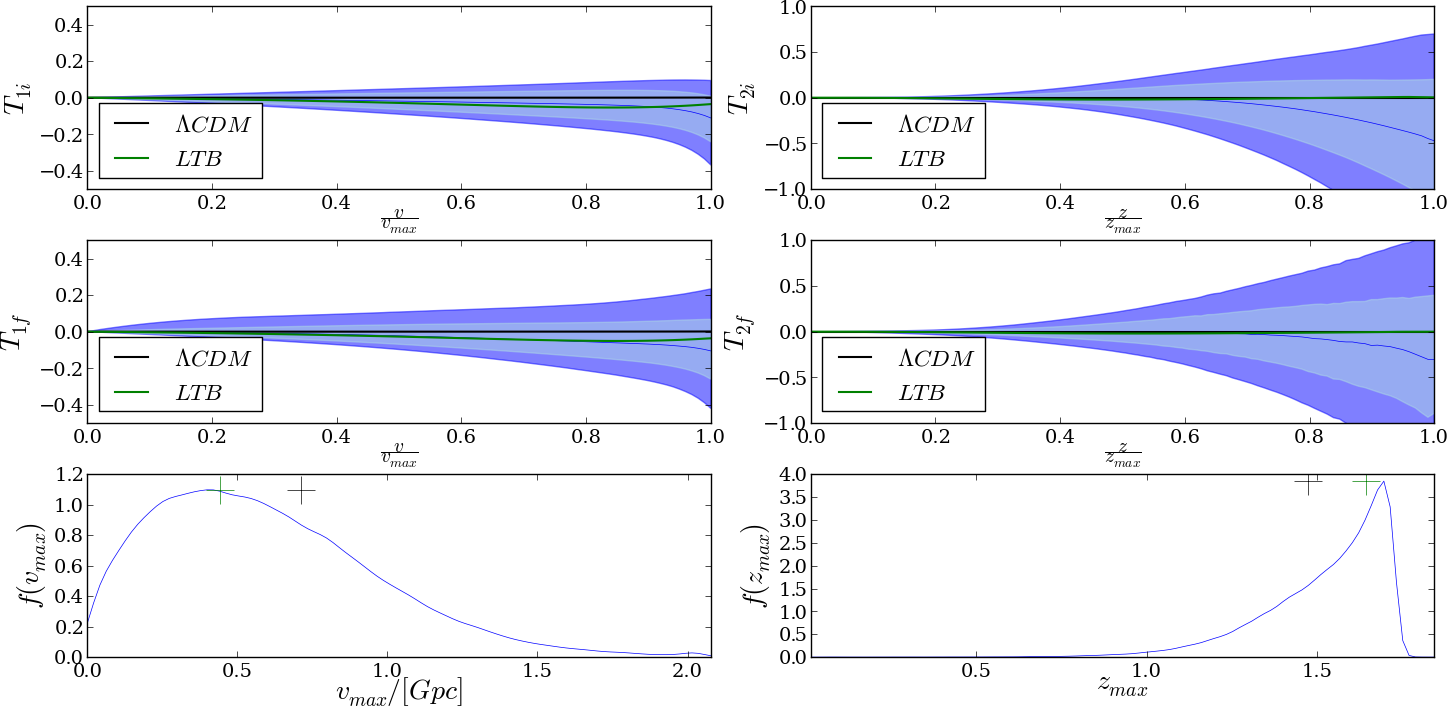}
\caption{Tests of the Copernican principle. Top Left: Posterior distribution of $T_1$ on the PLC0 marginalised over $t_0$.  Middle Left: Posterior distribution of $T_1$ on the PLCF. Bottom Left: The maximum extent of $v$ on the PLCF. The black cross is the background $\Lambda$CDM value. Top Right: Posterior distribution of $T_2$ on the PLC0 marginalised over $t_0$.  Middle Right: Posterior distribution of $T_2$ on the PLCF. Bottom Right: The maximum extent of $z$ on the PLCF. The black (green) crosses are the background $\Lambda$CDM (LTB) values. Clearly $\Lambda$CDM is compatible with the data. However we cannot yet rule out LTB model shown in green. \label{fig:RealTCP}}
\end{figure*}

\subsection{Simulations and verification}\label{sec:Sims}
To test our numerical implementation of the above algorithm we simulate data around a background $\Lambda$CDM model defined by
\begin{equation}
\Omega_{m0} = 0.3, \quad \Omega_{\Lambda 0} = 0.7, \quad H_0 = 70~\mbox{km}~\mbox{s}^{-1} \mbox{Mpc}^{-1}. 
\end{equation}
To simulate the data we have followed a similar procedure to that described in \cite{Bester:2013fya} except that we chose a simple power law relationship for how relative error scales with redshift viz.
\begin{equation}
\delta (z) = \delta_0(1+z)^\alpha.
\end{equation} 
Here $\delta_0$ is the relative error of the simulated data sets at $z = 0$. For both $D$ and $H$ we chose $\alpha = \frac{1}{3}$ with a relative error at the origin of 5\% and 10\% respectively. For the mock density data we chose $\alpha = 4$ with a relative error at the origin of 40\% and error bars centred on $\bar{\rho}$. We simulate a total of 500 data points for $D$ and 50 for both $H$ and $\rho$ with redshift values drawn from a uniform distribution between $z = 0$ and $z=2.0$. Finally, as explained in section~\ref{sec:Priors}, we use a Gaussian prior for $\Lambda$ with mean corresponding to the background $\Lambda$CDM value and an uncertainty of 2\%.\\
We then run these data through \ref{Alg1}. To make sure that the distributions have converged we run a number of these processes in parallel, each with a 10\% burn in period. The first process draws 1000 samples and each subsequent process doubles the number of samples generated by the last. We keep spawning new processes until the posterior distributions stops changing. Typical acceptance rates vary between 25\% and 35\%.\\ 
The results are summarised in Figures~\ref{fig:SimPLC0}-\ref{fig:Simtslice}. In all these figures the black crosses indicate the background $\Lambda$CDM values. Contours are found by constructing the empirical distribution function at each point in the domain. In all these figures the blue line is the median, the lightblue (dark blue) shaded area is the 1-$\sigma$ (2-$\sigma$) contours as they contain 68\% (95\%) of all realisations. As can clearly be seen from the figures the background model always falls within at least the 2-$\sigma$ contours of the reconstructed functions but are more often confined to the 1-$\sigma$ contours. It is especially reassuring that neither $T_1$ nor $T_2$ show a statistically significant departure from zero in Figure~\ref{fig:SimTCP}. Also note that the most probable density profile in Figure~\ref{fig:Simtslice} is very nearly flat but that void profiles do not seem to be ruled out. Figure~\ref{fig:Simtslice} also shows that the metric functions $X$ and $R$, although they exhibit significant uncertainty, both agree with what is expected from a flat $\Lambda$CDM model i.e. that $X$ is constant and $R$ a linear function of the comoving distance $r$.\\
There seems to be a very good agreement between the reconstructed functions and the background $\Lambda$CDM values. This confirms that our numerical implementation of \ref{Alg1} performs as expected.       

\begin{figure*}
\includegraphics[width=1.0\textwidth]{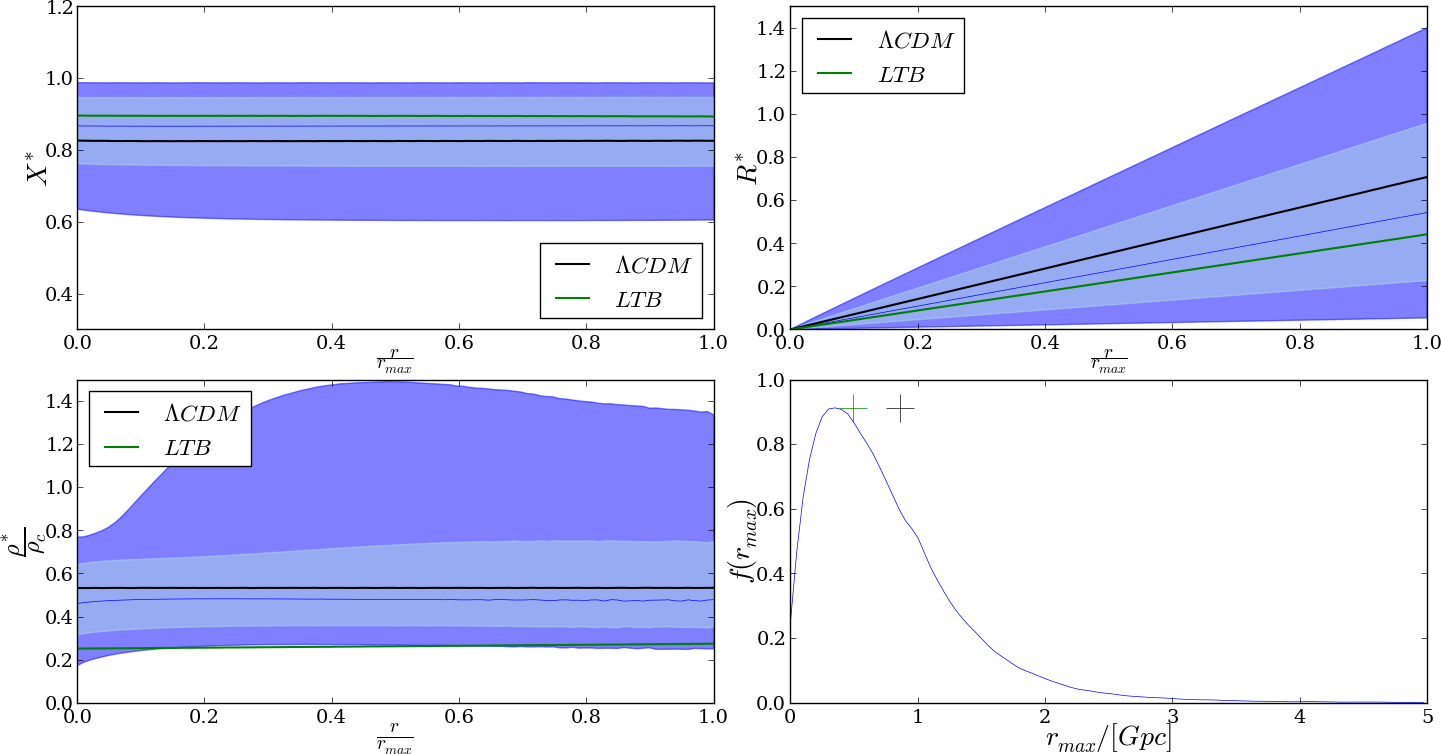}
\caption{Posterior distributions on a constant time slice $\Sigma_t$ at $t^* = 10.5$ Gyr for real data. Top Left: The metric function $X(t^*,r^*)$. This function is compatible with what is expected in $\Lambda$CDM since it is very nearly constant. However, because of the small volume probed by the current observations, we see that the LTB model (shown in green) seems to follow the same trend. The data would have to probe a much larger volume for the departure from this trend to be visible to the naked eye. Bottom Left: The density on $\Sigma_t$ plotted in units of the critical density $\rho_c$ of the background $\Lambda$CDM model used for the simulated data. Once again the most probable density profile is flat. However void profiles are not ruled out by the data. The density profile of the LTB model shown for comparison correspondes to that of a very shallow void. The overall value of this density is significantly smaller than that of $\Lambda$CDM and is almost ruled out at the 2-$\sigma$ confidence level. Top Right: The metric function $R(t^*,r^*)$ is very nearly a linear function of the comoving distance $r$ and is thus compatible with FLRW. However the same remarks as those regarding $X^*$ apply here as well. Bottom Right: The maximum extent of the radial coordinate $r^*$ on $\Sigma_t$. The black (green) cross is again the background $\Lambda$CDM (LTB) value.\label{fig:Realtslice}}
\end{figure*}

\section{Results and discussion}\label{sec:results}
The result of running currently available data through \ref{Alg1} are summarised in Figures~\ref{fig:RealPLC0}-\ref{fig:Realtslice}. For comparison we include a typical LTB model in the figures. The LTB model we use is the constrained LTB model of \cite{GarciaBellido2008nz} with parameters chosen to closely mimic the $\Lambda$CDM $D(z)$ relation. We are able to go out to a redshift of $z = 1.75$ as that is the maximum redshift for which we have $H(z)$ data. This places constraints on $D(z)$ beyond the maximum redshift of $z = 1.41$ in the Union 2.1 sample. Note that we use the same prior over $\rho(z)$ as for the simulated data. As explained in section~\ref{sec:Priors} we use a flat prior for $\Lambda$. These results confirm that the data are compatible with $\Lambda$CDM. Moreover we find the value of $H_0 = 69.76 \substack{+0.74 \\ -0.71} ~ \mbox{km} \mbox{s}^{-1} \mbox{Mpc}^{-1}$ at the 2-$\sigma$ confidence level. However the tight constraints on $H_0$ are artificial and result from not marginalising over its value when converting distance modulus to angular diameter distance data. \\
The similarity of the reconstructed distributions of $T_2$ on the initial and final PLC's, for both simulated and real data, suggests that it does not change by much during the evolution of the universe, at least for the class of models considered in this work. Furthermore this quantity seems to be quite insensitive to the value of $\Lambda$. It may therefore be able to shed some light on the Copernican principle even with our current uncertainty about this elusive parameter. It should be noted that the constraining power of angular diameter distance and expansion rate data diminishes at high redshifts. Density data at high $z$ should significantly constrain $T_2$. However, given that realistic models of the universe (i.e. FLRW + perturbations) should have a value of $T_2$ as low as $|T_2| \sim 10^{-5}$ \cite{Maartens:2011ITUH}, it remains to be seen whether density data from upcoming surveys will be sufficiently accurate to test the Copernican principle. A more detailed investigation is in preparation and will be released at a later stage. \\
The value of $\Lambda$ seems to have a greater impact on the evolution of the quantity $T_1$. However the reduced uncertainty of $T_1$ as compared to $T_2$ at high redshifts suggests that $T_1$ might be more appropriate to test the Copernican principle on large scales. Currently available data does not place very stringent constraints on the relative difference between $H_\perp$ and $H$. On the PLC0 these two quantities remain, at the 2-$\sigma$ confidence level, within 10\% of each other up to $z \approx 0.15$ but can differ by as much as 30\%-40\% at $z \approx 1.75$. As can be seen from Figure~\ref{fig:RealTCP} these constraints quickly degrade on the PLCF. Figure~\ref{fig:SimTCP} shows that the constraints are significantly better for our simulated data. Since the uncertainties in $D(z), H(z)$ and $\rho(z)$ are very similar for the simulated and real data it is the uncertainty in $\Lambda$ that degrades the constraints on $T_1$. Redshift drift data should therefore contribute significantly to tightening the constraints on $T_1$. This is another avenue that we are currently investigating, further details will be released at a later stage. \\
Neither $T_1$ nor $T_2$ are able to reject the void LTB model used for comparison. In fact the LTB model lies comfortably inside the 1-$\sigma$ contours of the current data. The observational signature of this model on the quantity $T_2$ is especially small. We found that using $H = H_\perp$ in \eqref{curvetest} results in a larger deviation of $T_2$ from zero. However the constraints from the data similarly degrade under this substitution so we do not gain anything from it. The inability of these tests to reject the LTB model suggests that much better data are required to test the Copernican principle.\\ 
Finally in Figure~\ref{fig:Realtslice} we show the distributions of the metric functions $X$ and $R$, as well as the distribution of $\rho$, on $\Sigma_t$. We have again included the LTB model for comparison. Both metric functions seem to follow trends typical of a flat FLRW cosmology i.e. that $X$ is constant and $R$ is a linear function of $r^*$. Although there is significant uncertainty in the values of these quantities it is not their values that we are after. We are really more interested to find out whether these functions can be considered as separable functions of $t$ and $r$. If they can it would be a nail in the coffin for radially inhomogeneous models since any spherically symmetric universe with separable metric components has to in fact be maximally symmetric and therefore homogeneous. Unfortunately it is very difficult to say with statistical certainty whether this is indeed the case. In fact the LTB metric components shown in these figures seem to follow the same trends. This is because of the relatively small volume that is accessible through observations at $t = 10.5$ Gyr. From the figure we see that this volume is less than 1 Gpc$^{3}$ whereas the full width at half maximum height of the void used to generate the LTB model is approximately 1.5 Gpc at $t=t_0$. The near flatness of the LTB density profile shows that the void has be very shallow at early times. However we also see that the LTB model's density would have been considerably smaller in the past so it might be possible to rule out certain classes of LTB models based on the allowed values of the density. Unfortunately we cannot draw any strong conclusions from this fact alone.  \\      
We have presented an algorithm that can in principle confront $\Lambda$LTB models with arbitrary data. Moreover the functional degrees of freedom of the $\Lambda$LTB model are fixed using directly observable quantities that are free from any gauge effects. We have further illustrated how $\Lambda$, the one free parameter of the model, can be constrained directly from observations of the redshift drift which is in principle completely model independent. As such we have presented a general framework for testing the Copernican principle both with currently available data and data from upcoming surveys. We have intentionally refrained from incorporating CMB data. The main reason for this is that the dynamics of non-comoving dust and radiation fluids in spherical symmetry is poorly understood \cite{2013JCAP...10..010L}. Self consistently incorporating radiation into the observational cosmology programme is highly non-trivial because we do not want to presuppose that the early universe is homogeneous. Instead the ideal is to model late time processes as accurately as possible and constrain the geometry of the universe based on data gathered with these models. Such an approach would be complementary to that employed in \cite{Redlich:2014gga} and \cite{2014arXiv1409.3831Z} for example. Both approaches should ultimately arrive at the same conclusions. \\
Furthermore in section~\ref{sec:inference} we suggested a way to incorporate data that are not entirely model independent. This is obviously not the ideal way to test the Copernican principle. However the considerable difficulty involved in getting model independent data might necessitate this kind of approach for some time to come. We should also keep in mind that this framework relies on the validity of a number of additional assumptions. As a result the tests we perform do not only test the Copernican principle but also the compatibility of the data with these assumptions. We should therefore strive to confront the data with as many consistency relations as possible. It would seem, at least within the framework presented in this paper, that there is some work to be done before we can ultimately confirm or refute the validity of the Copernican principle.

\section*{Acknowledgements}
We would like to thank Rhodes university and the Rhodes mathematics department for the use of the GetaFix cluster. We also thank Petrus van der Walt, Chris Clarkson, Prina Patel, David Bacon, Bruce Bassett, Charles Hellaby, Anthony Walters and Sean February for their useful contributions during discussions. A further thank you is extended to  Petrus van der Walt for making his CIVP code available. JL and NTB are supported by the National Research Foundation (South Africa). The financial assistance of the South African Square Kilometre Array project (SKA SA) towards HLB's research is hereby acknowledged. Opinions expressed and conclusions arrived at are those of the author and are not necessarily to be attributed to the SKA SA (www.ska.ac.za).

\bibliographystyle{mnras}
\bibliography{Towards_the_geometry_of_the_universe_from_data}

\begin{thebibliography}{}
\makeatletter
\relax
\def\mn@urlcharsother{\let\do\@makeother \do\$\do\&\do\#\do\^\do\_\do\%\do\~}
\def\mn@doi{\begingroup\mn@urlcharsother \@ifnextchar [ {\mn@doi@}
  {\mn@doi@[]}}
\def\mn@doi@[#1]#2{\def\@tempa{#1}\ifx\@tempa\@empty \href
  {http://dx.doi.org/#2} {doi:#2}\else \href {http://dx.doi.org/#2} {#1}\fi
  \endgroup}
\def\mn@eprint#1#2{\mn@eprint@#1:#2::\@nil}
\def\mn@eprint@arXiv#1{\href {http://arxiv.org/abs/#1} {{\tt arXiv:#1}}}
\def\mn@eprint@dblp#1{\href {http://dblp.uni-trier.de/rec/bibtex/#1.xml}
  {dblp:#1}}
\def\mn@eprint@#1:#2:#3:#4\@nil{\def\@tempa {#1}\def\@tempb {#2}\def\@tempc
  {#3}\ifx \@tempc \@empty \let \@tempc \@tempb \let \@tempb \@tempa \fi \ifx
  \@tempb \@empty \def\@tempb {arXiv}\fi \@ifundefined
  {mn@eprint@\@tempb}{\@tempb:\@tempc}{\expandafter \expandafter \csname
  mn@eprint@\@tempb\endcsname \expandafter{\@tempc}}}

\bibitem[\protect\citeauthoryear{Ade et~al.}{Ade et~al.}{2014}]{Ade:2013sjv}
Ade P.,  et~al., 2014, \mn@doi [Astron.Astrophys.]
  {10.1051/0004-6361/201321529}, 571, A1

\bibitem[\protect\citeauthoryear{{Allen}, {Evrard}  \& {Mantz}}{{Allen}
  et~al.}{2011}]{2011ARA&A..49..409A}
{Allen} S.~W.,  {Evrard} A.~E.,   {Mantz} A.~B.,  2011, \mn@doi [\araa]
  {10.1146/annurev-astro-081710-102514}, \href
  {http://adsabs.harvard.edu/abs/2011ARA%26A..49..409A} {49, 409}

\bibitem[\protect\citeauthoryear{{Amanullah} et~al.,}{{Amanullah}
  et~al.}{2010}]{2010ApJ...716..712A}
{Amanullah} R.,  et~al., 2010, \mn@doi [\apj] {10.1088/0004-637X/716/1/712},
  \href {http://adsabs.harvard.edu/abs/2010ApJ...716..712A} {716, 712}

\bibitem[\protect\citeauthoryear{{Amendola} et~al.,}{{Amendola}
  et~al.}{2013}]{2013LRR....16....6A}
{Amendola} L.,  et~al., 2013, \mn@doi [Living Reviews in Relativity]
  {10.12942/lrr-2013-6}, \href
  {http://adsabs.harvard.edu/abs/2013LRR....16....6A} {16, 6}

\bibitem[\protect\citeauthoryear{{Ara{\'u}jo} \& {Stoeger}}{{Ara{\'u}jo} \&
  {Stoeger}}{1999}]{araujo00}
{Ara{\'u}jo} M.~E.,  {Stoeger} W.~R.,  1999, \mn@doi [Phys. Rev. D]
  {10.1103/PhysRevD.60.104020}, \href
  {http://esoads.eso.org/abs/1999PhRvD..60j4020A} {60, 104020}

\bibitem[\protect\citeauthoryear{{Ara{\'u}jo} \& {Stoeger}}{{Ara{\'u}jo} \&
  {Stoeger}}{2009}]{araujo09.1}
{Ara{\'u}jo} M.~E.,  {Stoeger} W.~R.,  2009, \mn@doi [Mon. Not. Roy. Astr.
  Soc.] {10.1111/j.1365-2966.2008.14321.x}, \href
  {http://esoads.eso.org/abs/2009MNRAS.394..438A} {394, 438}

\bibitem[\protect\citeauthoryear{{Ara{\'u}jo} \& {Stoeger}}{{Ara{\'u}jo} \&
  {Stoeger}}{2010}]{araujo10}
{Ara{\'u}jo} M.~E.,  {Stoeger} W.~R.,  2010, \mn@doi [Phys. Rev. D]
  {10.1103/PhysRevD.82.123513}, \href
  {http://esoads.eso.org/abs/2010PhRvD..82l3513A} {82, 123513}

\bibitem[\protect\citeauthoryear{{Ara{\'u}jo} \& {Stoeger}}{{Ara{\'u}jo} \&
  {Stoeger}}{2011}]{araujo11}
{Ara{\'u}jo} M.~E.,  {Stoeger} W.~R.,  2011, \mn@doi [JCAP]
  {10.1088/1475-7516/2011/07/029}, \href
  {http://esoads.eso.org/abs/2011JCAP...07..029A} {7, 29}

\bibitem[\protect\citeauthoryear{{{Ara{\'u}jo}, M.E. and Stoeger,
  W.R.}}{{{Ara{\'u}jo}, M.E. and Stoeger, W.R.}}{2009}]{Araujo2009zh}
{{Ara{\'u}jo}, M.E. and Stoeger, W.R.} 2009, \mn@doi [Phys.Rev.]
  {10.1103/PhysRevD.80.123517, 10.1103/PhysRevD.81.049903}, D80, 123517

\bibitem[\protect\citeauthoryear{Audren}{Audren}{2014}]{Audren11102014}
Audren B.,  2014, \mn@doi [Monthly Notices of the Royal Astronomical Society]
  {10.1093/mnras/stu1457}, 444, 827

\bibitem[\protect\citeauthoryear{{Balbi} \& {Quercellini}}{{Balbi} \&
  {Quercellini}}{2007}]{2007MNRAS.382.1623B}
{Balbi} A.,  {Quercellini} C.,  2007, \mn@doi [MNRAS]
  {10.1111/j.1365-2966.2007.12407.x}, \href
  {http://adsabs.harvard.edu/abs/2007MNRAS.382.1623B} {382, 1623}

\bibitem[\protect\citeauthoryear{Bartelmann \& Schneider}{Bartelmann \&
  Schneider}{2001}]{Bartelmann:1999yn}
Bartelmann M.,  Schneider P.,  2001, \mn@doi [Phys.Rept.]
  {10.1016/S0370-1573(00)00082-X}, 340, 291

\bibitem[\protect\citeauthoryear{{Bassett} \& {Hlozek}}{{Bassett} \&
  {Hlozek}}{2010}]{2010deot.book..246B}
{Bassett} B.,  {Hlozek} R.,  2010, {Baryon acoustic oscillations}.
Cambridge Univ. Press, p.~246

\bibitem[\protect\citeauthoryear{Bester, Larena, van~der Walt  \&
  Bishop}{Bester et~al.}{2014}]{Bester:2013fya}
Bester H.~L.,  Larena J.,  van~der Walt P.~J.,   Bishop N.~T.,  2014, \mn@doi
  [JCAP] {10.1088/1475-7516/2014/02/009}, 1402, 009

\bibitem[\protect\citeauthoryear{Buchert \& Räsänen}{Buchert \&
  Räsänen}{2012}]{Buchert:2011sx}
Buchert T.,  Räsänen S.,  2012, \mn@doi [Ann.Rev.Nucl.Part.Sci.]
  {10.1146/annurev.nucl.012809.104435}, 62, 57

\bibitem[\protect\citeauthoryear{{Busti}, {Clarkson}  \& {Seikel}}{{Busti}
  et~al.}{2014a}]{2014IAUS..306...25B}
{Busti} V.~C.,  {Clarkson} C.,   {Seikel} M.,  2014a, in IAU Symposium. pp
  25--27 (\mn@eprint {arXiv} {1407.5227}), \mn@doi{10.1017/S1743921314013751}

\bibitem[\protect\citeauthoryear{{Busti}, {Clarkson}  \& {Seikel}}{{Busti}
  et~al.}{2014b}]{2014MNRAS.441L..11B}
{Busti} V.~C.,  {Clarkson} C.,   {Seikel} M.,  2014b, \mn@doi [MNRAS]
  {10.1093/mnrasl/slu035}, \href
  {http://adsabs.harvard.edu/abs/2014MNRAS.441L..11B} {441, L11}

\bibitem[\protect\citeauthoryear{Clarkson, Bassett  \& Lu}{Clarkson
  et~al.}{2008}]{Clarkson:2007pz}
Clarkson C.,  Bassett B.,   Lu T. H.-C.,  2008, \mn@doi [Phys.Rev.Lett.]
  {10.1103/PhysRevLett.101.011301}, 101, 011301

\bibitem[\protect\citeauthoryear{Clarkson, Ellis, Larena  \& Umeh}{Clarkson
  et~al.}{2011}]{Clarkson:2011zq}
Clarkson C.,  Ellis G.,  Larena J.,   Umeh O.,  2011, \mn@doi [Rept.Prog.Phys.]
  {10.1088/0034-4885/74/11/112901}, 74, 112901

\bibitem[\protect\citeauthoryear{Corasaniti, Huterer  \& Melchiorri}{Corasaniti
  et~al.}{2007}]{Corasaniti:2007bg}
Corasaniti P.-S.,  Huterer D.,   Melchiorri A.,  2007, \mn@doi [Phys.Rev.]
  {10.1103/PhysRevD.75.062001}, D75, 062001

\bibitem[\protect\citeauthoryear{Cotter, Roberts, Stuart  \& White}{Cotter
  et~al.}{2013}]{cotter2013}
Cotter S.~L.,  Roberts G.~O.,  Stuart A.~M.,   White D.,  2013, \mn@doi
  [Statist. Sci.] {10.1214/13-STS421}, 28, 424

\bibitem[\protect\citeauthoryear{{Ellis}, {Nel}, {Maartens}, {Stoeger}  \&
  {Whitman}}{{Ellis} et~al.}{1985}]{1985PhR...124..315E}
{Ellis} G.~F.~R.,  {Nel} S.~D.,  {Maartens} R.,  {Stoeger} W.~R.,   {Whitman}
  A.~P.,  1985, \mn@doi [\physrep] {10.1016/0370-1573(85)90030-4}, \href
  {http://adsabs.harvard.edu/abs/1985PhR...124..315E} {124, 315}

\bibitem[\protect\citeauthoryear{Garcia-Bellido \& Haugboelle}{Garcia-Bellido
  \& Haugboelle}{2008}]{GarciaBellido2008nz}
Garcia-Bellido J.,  Haugboelle T.,  2008, \mn@doi [JCAP]
  {10.1088/1475-7516/2008/04/003}, 0804, 003

\bibitem[\protect\citeauthoryear{Gardner}{Gardner}{1998}]{Gardner:1998ae}
Gardner J.~P.,  1998, \mn@doi [Publ.Astron.Soc.Pac.] {10.1086/316141}, 110, 291

\bibitem[\protect\citeauthoryear{Geng, Li, Zhang  \& Zhang}{Geng
  et~al.}{2015}]{Geng:2015ara}
Geng J.-J.,  Li Y.-H.,  Zhang J.-F.,   Zhang X.,  2015, \mn@doi [Eur. Phys. J.]
  {10.1140/epjc/s10052-015-3581-8}, C75, 356

\bibitem[\protect\citeauthoryear{Hellaby}{Hellaby}{2001}]{Hellaby:2000ef}
Hellaby C.,  2001, \mn@doi [Astron.Astrophys.] {10.1051/0004-6361:20010172},
  372, 357

\bibitem[\protect\citeauthoryear{Hellaby}{Hellaby}{2013}]{Hellaby:2013ts}
Hellaby C.,  2013, Proceedings of the 27th Texas Symposium on Relativistic
  Astrophysics

\bibitem[\protect\citeauthoryear{Humphreys, Maartens  \& Matravers}{Humphreys
  et~al.}{2012}]{Humphreys:1998tc}
Humphreys N.~P.,  Maartens R.,   Matravers D.~R.,  2012, \mn@doi
  [Gen.Rel.Grav.] {10.1007/s10714-012-1452-2}, 44, 3197

\bibitem[\protect\citeauthoryear{Krasi\ifmmode~\acute{n}\else
  \'{n}\fi{}ski}{Krasi\ifmmode~\acute{n}\else
  \'{n}\fi{}ski}{2014}]{PhysRevD.90.064021}
Krasi\ifmmode~\acute{n}\else \'{n}\fi{}ski A.,  2014, \mn@doi [Phys. Rev. D]
  {10.1103/PhysRevD.90.064021}, 90, 064021

\bibitem[\protect\citeauthoryear{{Lim}, {Regis}  \& {Clarkson}}{{Lim}
  et~al.}{2013}]{2013JCAP...10..010L}
{Lim} W.~C.,  {Regis} M.,   {Clarkson} C.,  2013, \mn@doi [\jcap]
  {10.1088/1475-7516/2013/10/010}, \href
  {http://adsabs.harvard.edu/abs/2013JCAP...10..010L} {10, 10}

\bibitem[\protect\citeauthoryear{Loeb}{Loeb}{1998}]{Loeb:1998bu}
Loeb A.,  1998, \mn@doi [Astrophys.J.] {10.1086/311375}, 499, L111

\bibitem[\protect\citeauthoryear{{Maartens}}{{Maartens}}{2011}]{Maartens:2011ITUH}
{Maartens} R.,  2011, \mn@doi [Royal Society of London Philosophical
  Transactions Series A] {10.1098/rsta.2011.0289}, \href
  {http://adsabs.harvard.edu/abs/2011RSPTA.369.5115M} {369, 5115}

\bibitem[\protect\citeauthoryear{{Maartens}, {Humphreys}, {Matravers}  \&
  {Stoeger}}{{Maartens} et~al.}{1996}]{Maartens96}
{Maartens} R.,  {Humphreys} N.~P.,  {Matravers} D.~R.,   {Stoeger} W.~R.,
  1996, \mn@doi [Class. Quantum Grav.] {10.1088/0264-9381/13/2/013}, \href
  {http://esoads.eso.org/abs/1996CQGra..13..253M} {13, 253}

\bibitem[\protect\citeauthoryear{Maartens, Abdalla, Jarvis  \& Santos}{Maartens
  et~al.}{2015}]{Maartens:2015mra}
Maartens R.,  Abdalla F.~B.,  Jarvis M.,   Santos M.~G.,  2015, PoS, AASKA14,
  016

\bibitem[\protect\citeauthoryear{Moresco, Verde, Pozzetti, Jimenez  \&
  Cimatti}{Moresco et~al.}{2012}]{Moresco:2012by}
Moresco M.,  Verde L.,  Pozzetti L.,  Jimenez R.,   Cimatti A.,  2012, \mn@doi
  [JCAP] {10.1088/1475-7516/2012/07/053}, 1207, 053

\bibitem[\protect\citeauthoryear{Percival, Samushia, Ross, Shapiro  \&
  Raccanelli}{Percival et~al.}{2011}]{redspacedistotions2011}
Percival W.~J.,  Samushia L.,  Ross A.~J.,  Shapiro C.,   Raccanelli A.,  2011,
  \mn@doi [Philosophical Transactions of the Royal Society of London A:
  Mathematical, Physical and Engineering Sciences] {10.1098/rsta.2011.0370},
  369, 5058

\bibitem[\protect\citeauthoryear{Rasmussen \& Williams}{Rasmussen \&
  Williams}{2006}]{rasmussen2006gaussian}
Rasmussen C.,  Williams C.,  2006, Gaussian Processes for Machine Learning.
Adaptative computation and machine learning series, University Press Group
  Limited, \url {http://books.google.ca/books?id=vWtwQgAACAAJ}

\bibitem[\protect\citeauthoryear{Redlich, Bolejko, Meyer, Lewis  \&
  Bartelmann}{Redlich et~al.}{2014}]{Redlich:2014gga}
Redlich M.,  Bolejko K.,  Meyer S.,  Lewis G.~F.,   Bartelmann M.,  2014,
  \mn@doi [Astron.Astrophys.] {10.1051/0004-6361/201424553}, 570, A63

\bibitem[\protect\citeauthoryear{S\'anchez \& the Des~collaboration}{S\'anchez
  \& the Des~collaboration}{2010}]{1742-6596-259-1-012080}
S\'anchez E.,  the Des~collaboration 2010, Journal of Physics: Conference
  Series, 259, 012080

\bibitem[\protect\citeauthoryear{{Sandage}}{{Sandage}}{1962}]{1962ApJ...136..319S}
{Sandage} A.,  1962, \mn@doi [\apj] {10.1086/147385}, \href
  {http://adsabs.harvard.edu/abs/1962ApJ...136..319S} {136, 319}

\bibitem[\protect\citeauthoryear{Sapone, Majerotto  \& Nesseris}{Sapone
  et~al.}{2014}]{Sapone:2014nna}
Sapone D.,  Majerotto E.,   Nesseris S.,  2014, \mn@doi [Phys.Rev.]
  {10.1103/PhysRevD.90.023012}, D90, 023012

\bibitem[\protect\citeauthoryear{Seikel, Clarkson  \& Smith}{Seikel
  et~al.}{2012}]{Seikel:2012uu}
Seikel M.,  Clarkson C.,   Smith M.,  2012, \mn@doi [JCAP]
  {10.1088/1475-7516/2012/06/036}, 1206, 036

\bibitem[\protect\citeauthoryear{{Shafieloo}, {Kim}  \& {Linder}}{{Shafieloo}
  et~al.}{2012}]{2012PhRvD..85l3530S}
{Shafieloo} A.,  {Kim} A.~G.,   {Linder} E.~V.,  2012, \mn@doi [Phys. Rev. D]
  {10.1103/PhysRevD.85.123530}, \href
  {http://adsabs.harvard.edu/abs/2012PhRvD..85l3530S} {85, 123530}

\bibitem[\protect\citeauthoryear{{Stoeger}, {Nel}, {Maartens}  \&
  {Ellis}}{{Stoeger} et~al.}{1992a}]{Stoeger92.0}
{Stoeger} W.~R.,  {Nel} S.~D.,  {Maartens} R.,   {Ellis} G.~F.~R.,  1992a,
  \mn@doi [Class. Quantum Grav.] {10.1088/0264-9381/9/2/013}, \href
  {http://esoads.eso.org/abs/1992CQGra...9..493S} {9, 493}

\bibitem[\protect\citeauthoryear{{Stoeger}, {Ellis}  \& {Nel}}{{Stoeger}
  et~al.}{1992b}]{Stoeger92.1}
{Stoeger} W.~R.,  {Ellis} G.~F.~R.,   {Nel} S.~D.,  1992b, \mn@doi [Class.
  Quantum Grav.] {10.1088/0264-9381/9/2/014}, \href
  {http://esoads.eso.org/abs/1992CQGra...9..509S} {9, 509}

\bibitem[\protect\citeauthoryear{{Stoeger}, {Stanley}, {Nel}  \&
  {Ellis}}{{Stoeger} et~al.}{1992c}]{Stoeger92.2}
{Stoeger} W.~R.,  {Stanley} S.~J.,  {Nel} D.,   {Ellis} G.~F.~R.,  1992c,
  \mn@doi [Class. Quantum Grav.] {10.1088/0264-9381/9/7/007}, \href
  {http://esoads.eso.org/abs/1992CQGra...9.1711S} {9, 1711}

\bibitem[\protect\citeauthoryear{{Stoeger}, {Stanley}, {Nel}  \&
  {Ellis}}{{Stoeger} et~al.}{1992d}]{Stoeger92.3}
{Stoeger} W.~R.,  {Stanley} S.~J.,  {Nel} D.,   {Ellis} G.~F.~R.,  1992d,
  \mn@doi [Class. Quantum Grav.] {10.1088/0264-9381/9/7/008}, \href
  {http://esoads.eso.org/abs/1992CQGra...9.1725S} {9, 1725}

\bibitem[\protect\citeauthoryear{{Stoeger}, {Ellis}  \& {Xu}}{{Stoeger}
  et~al.}{1994}]{Stoeger94}
{Stoeger} W.~R.,  {Ellis} G.~F.~R.,   {Xu} C.,  1994, \mn@doi [Phys. Rev. D]
  {10.1103/PhysRevD.49.1845}, \href
  {http://esoads.eso.org/abs/1994PhRvD..49.1845S} {49, 1845}

\bibitem[\protect\citeauthoryear{{Stoeger}, {Araujo}  \& {Gebbie}}{{Stoeger}
  et~al.}{1997}]{Stoeger97}
{Stoeger} W.~R.,  {Araujo} M.~E.,   {Gebbie} T.,  1997, \mn@doi [Astrophys. J.]
  {10.1086/303633}, \href {http://esoads.eso.org/abs/1997ApJ...476..435S} {476,
  435}

\bibitem[\protect\citeauthoryear{Suzuki, Rubin, Lidman, Aldering, Amanullah
  et~al.}{Suzuki et~al.}{2012}]{Suzuki:2011hu}
Suzuki N.,  Rubin D.,  Lidman C.,  Aldering G.,  Amanullah R.,   et~al., 2012,
  \mn@doi [Astrophys.J.] {10.1088/0004-637X/746/1/85}, 746, 85

\bibitem[\protect\citeauthoryear{Uzan, Clarkson  \& Ellis}{Uzan
  et~al.}{2008}]{PhysRevLett.100.191303}
Uzan J.-P.,  Clarkson C.,   Ellis G. F.~R.,  2008, \mn@doi [Phys. Rev. Lett.]
  {10.1103/PhysRevLett.100.191303}, 100, 191303

\bibitem[\protect\citeauthoryear{Valkenburg}{Valkenburg}{2012}]{Valkenburg:2011tm}
Valkenburg W.,  2012, \mn@doi [Gen.Rel.Grav.] {10.1007/s10714-012-1405-9}, 44,
  2449

\bibitem[\protect\citeauthoryear{{Valkenburg}, {Marra}  \&
  {Clarkson}}{{Valkenburg} et~al.}{2014}]{2014MNRAS.438L...6V}
{Valkenburg} W.,  {Marra} V.,   {Clarkson} C.,  2014, \mn@doi [MNRAS]
  {10.1093/mnrasl/slt140}, \href
  {http://adsabs.harvard.edu/abs/2014MNRAS.438L...6V} {438, L6}

\bibitem[\protect\citeauthoryear{{Vonlanthen}, {R{\"a}s{\"a}nen}  \&
  {Durrer}}{{Vonlanthen} et~al.}{2010}]{2010JCAP...08..023V}
{Vonlanthen} M.,  {R{\"a}s{\"a}nen} S.,   {Durrer} R.,  2010, \mn@doi [\jcap]
  {10.1088/1475-7516/2010/08/023}, \href
  {http://adsabs.harvard.edu/abs/2010JCAP...08..023V} {8, 23}

\bibitem[\protect\citeauthoryear{Walters \& Hellaby}{Walters \&
  Hellaby}{2012}]{Walters:2012ns}
Walters A.,  Hellaby C.,  2012, \mn@doi [JCAP] {10.1088/1475-7516/2012/12/001},
  1212, 001

\bibitem[\protect\citeauthoryear{{Zibin} \& {Moss}}{{Zibin} \&
  {Moss}}{2014}]{2014arXiv1409.3831Z}
{Zibin} J.~P.,  {Moss} A.,  2014, preprint, \href
  {http://adsabs.harvard.edu/abs/2014arXiv1409.3831Z} {} (\mn@eprint {arXiv}
  {1409.3831})

\bibitem[\protect\citeauthoryear{{van der Walt} \& {Bishop}}{{van der Walt} \&
  {Bishop}}{2012}]{prd2012}
{van der Walt} P.~J.,  {Bishop} N.~T.,  2012, \mn@doi [Phys. Rev. D]
  {10.1103/PhysRevD.85.044016}, \href
  {http://esoads.eso.org/abs/2012PhRvD..85d4016V} {85, 044016}

\makeatother
\end{thebibliography}

\end{document}